\documentclass[journal, onecolumn]{IEEEtran}
\renewcommand{\baselinestretch}{2.0}

\ifCLASSINFOpdf
\else
\fi

\include{epsf}
\include{epsfrot}
\usepackage[dvips]{graphicx}
\usepackage{epsfig}
\usepackage{afterpage}
\usepackage{graphicx}
\usepackage[centertags]{amsmath}
\usepackage{amsfonts}
\usepackage{amssymb}
\usepackage{amsthm}
\usepackage{nccmath}
\usepackage{color}

\makeatletter
\newtheorem{remark}{Remark}

\def \be {\begin{eqnarray}}
\def \ee {\end{eqnarray}}
\def \bc {\begin{center}}
\def \ec {\end{center}}
\def\bfig { \begin{figure} }
\def\efig { \end{figure} }
\def\bit { \begin{itemize} }
\def\eit { \end{itemize} }
\def\benum { \begin{enumerate} }
\def\eenum { \end{enumerate} }

\begin{document}
%
\title{HARQ Buffer Management: An Information-Theoretic View}
%
%
%

\author{Wonju~Lee,~\IEEEmembership{Student~Member,~IEEE,}
        Osvaldo~Simeone,~\IEEEmembership{Senior~Member,~IEEE,}
        Joonhyuk~Kang,~\IEEEmembership{Member,~IEEE,}
        Sundeep~Rangan,~\IEEEmembership{Senior~Member,~IEEE},
        and Petar~Popovski,~\IEEEmembership{Senior~Member,~IEEE}
\thanks{W. Lee and J. Kang are with the Department of Electrical Engineering, Korea Advanced Institute of Science and Technology (KAIST), Daejeon,
305-701, South Korea (e-mail: wonjulee@kaist.ac.kr, jhkang@ee.kaist.ac.kr).}
\thanks{O. Simeone is with the Center for Wireless Communications and Signal Processing Research (CWCSPR),
Department of Electrical and Computer Engineering, New Jersey Institute of Technology (NJIT), Newark, NJ 07102, USA
(e-mail: osvaldo.simeone@njit.edu).}
\thanks{S. Rangan is with the New York University Wireless Center, Department of Electrical and Computer Engineering, Polytechnic Institute of New York University (NYU), Brooklyn, NY 11201, USA
(e-mail: srangan@poly.edu).}
\thanks{P. Popovski is with the Antennas, Propagation, and Radio Networking (APNET), Department of Electronic Systems, Aalborg University, Aalborg, 9220, Denmark
(e-mail: petarp@es.aau.dk).}}


%
%

\markboth{Journal of \LaTeX\ Class Files, February~2015}%
{Shell \MakeLowercase{\textit{et al.}}: Bare Demo of IEEEtran.cls for Journals}
%

\maketitle

\begin{abstract}
A key practical constraint on the design of Hybrid automatic repeat request (HARQ) schemes is the size of the on-chip buffer that is available at the receiver to store previously received packets.
In fact, in modern wireless standards such as LTE and LTE-A, the HARQ buffer size is one of the main drivers of the modem area and power consumption.
This has recently highlighted the importance of HARQ buffer management,
that is, of the use of buffer-aware transmission schemes and of advanced compression policies for the storage of received data.
This work investigates HARQ buffer management by leveraging information-theoretic achievability arguments based on random coding.
Specifically, standard HARQ schemes, namely Type-I, Chase Combining and Incremental Redundancy, are first studied under the assumption of a finite-capacity
HARQ buffer by considering both coded modulation, via Gaussian signaling, and Bit Interleaved Coded Modulation (BICM).
The analysis sheds light on the impact of different compression strategies, namely the conventional compression log-likelihood ratios and the direct digitization of baseband signals, on the throughput.
Then, coding strategies based on layered modulation and optimized coding blocklength are investigated, highlighting the benefits of HARQ buffer-aware transmission schemes.
The optimization of baseband compression for multiple-antenna links is also studied, demonstrating the optimality of a transform coding approach.
\end{abstract}
%

%
\IEEEpeerreviewmaketitle

\section{Introduction}
Hybrid automatic repeat request (HARQ) is an integral part of modern wireless communication standards such as LTE and LTE-A \cite{Andrews,sesia2009lte}.
HARQ enables reliable communication over time-varying fading channel by leveraging both forward error-correcting coding at the physical layer
and automatic retransmissions at the data link/medium access layer based on binary ACK/NACK feedback on the reverse link.
With HARQ, the receiver can store previously received packets for joint processing with the last received packet in order to enhance the decoding reliability.
Three HARQ mechanisms are conventionally used, namely HARQ Type I (HARQ-TI), HARQ Chase Combining (HARQ-CC), and HARQ Incremental Redundancy (HARQ-IR) (see, e.g., \cite{Andrews}-\cite{Chase}).

One of the key challenges in implementing HARQ is the need to store data from previously received packets on chip.
In LTE and LTE-A, the HARQ buffer is in fact one of the main drivers of the overall modem area and power consumption, as well as a key determinant of the User Equipment (UE) category level \cite{sesia2009lte,Bai}.
Placing the HARQ buffer off chip can also be challenging due to the large bandwidth requirements on the external memory interface.
These problems are expected to become even more severe for the next-generation systems, e.g., based on mmWave technology \cite{RanRapE:14,Rappaport}, due to the larger bandwidth and transmission rates.

The limitations in the HARQ buffer size dictated by the modem area and power consumption make the use of buffer-aware transmission strategies and
of advanced compression\footnote{In this paper, compression is meant to include also the step of quantization.} policies for the storage of received data of critical importance for the feasibility of HARQ in modern wireless standards \cite{Bai,Danieli}.
An example of the former is limited buffer rate matching in LTE \cite{sesia2009lte} and an instance of the latter is the vector quantization scheme proposed in \cite{Danieli} to store the log-likelihood ratios (LLRs) of the coded bits for the previously received packets.
We refer to transmit- and receive-side mechanisms meant to cope with HARQ buffer limitations as HARQ buffer management.

Previous theoretical work on HARQ has assumed unrestricted HARQ buffers to be available at the receivers
or has imposed limits on the number of packets that can be stored (see, e.g., \cite{Costello,Caire} and references therein).
In this paper, instead, we assume a generic capacity constraint for the HARQ buffer in terms of number of bits, and we aim at addressing the following main questions:
(\textit{i}) How is the relative performance of standard HARQ schemes, namely HARQ-TI, HARQ-CC and HARQ-IR, affected by the amount of available HARQ buffer capacity?
(\textit{ii}) Are there more efficient alternatives to the conventional approach of representing buffered packets at the receiver by quantizing the LLRs of the coded bits (see \cite{Bai,Danieli})?
(\textit{iii}) What is the impact of buffer-aware transmission strategies such as layered modulation and rate matching?
(\textit{iv}) What new opportunities and challenges arise in the design of HARQ buffer management for multiple-antenna (MIMO) links?

This works makes some steps towards answering these questions by leveraging information-theoretic achievability arguments based on random coding.
Our contributions are as follows.
\begin{itemize}
\item We study a baseline system that uses an ideal coded modulation scheme via Gaussian signaling at the transmitter and compression of the previously received packets at the baseband level
with the aim of assessing the impact of a finite HARQ buffer on the throughput of HARQ-TI, HARQ-CC and HARQ-IR (Sec. \ref{ch:HARQ_BBcomp}).
\item We investigate the more complex case of a link employing Bit Interleaved Coded Modulation (BICM) \cite{Caire2} and study the performance with both baseband compression and the more conventional LLR compression of the previously received packets (Sec. \ref{ch:BICM_Comp}).
The goal of the analysis is to address the possible suboptimality of the conventional approach of quantizing LLRs for storage in the HARQ buffer.
\item We study the potential benefits of buffer-aware transmission strategies based on layered transmission \cite{Steiner}, whereby the rates of the transmission layers are adopted to the HARQ buffer size (Sec. \ref{ch:LayeredCoding}).
\item We study the design of baseband compression for a link with multiple-antennas and show the optimality of a compression strategy based on transform coding (Sec. \ref{ch:MIMO}).
\item We analyze the impact of the selection of the transmission blocklength as a function of the HARQ buffer size (Sec. \ref{ch:F_Blocklength}).
This analysis complements the study in \cite{Devassy}, which assumed no buffer limitations.
\end{itemize}

Finally, Sec. \ref{ch:simulation} presents numerical results and Sec. \ref{ch:conclusion} offers with some concluding remarks\footnote{The content of Sec. \ref{ch:HARQ_BBcomp} and Sec. \ref{ch:BICM_Comp} was partially presented in \cite{WLee}.}.

\textit{Notation}: $(\cdot)^{*}$ denotes the complex transpose; $E[\cdot]$ is the expectation operator; information-theoretic quantities such as mutual information are defined as in \cite{Cover}.

\section{System Model and Performance Criteria} \label{ch:systemmodel}
\begin{figure}[t]
\begin{center}
\centering \epsfig{file=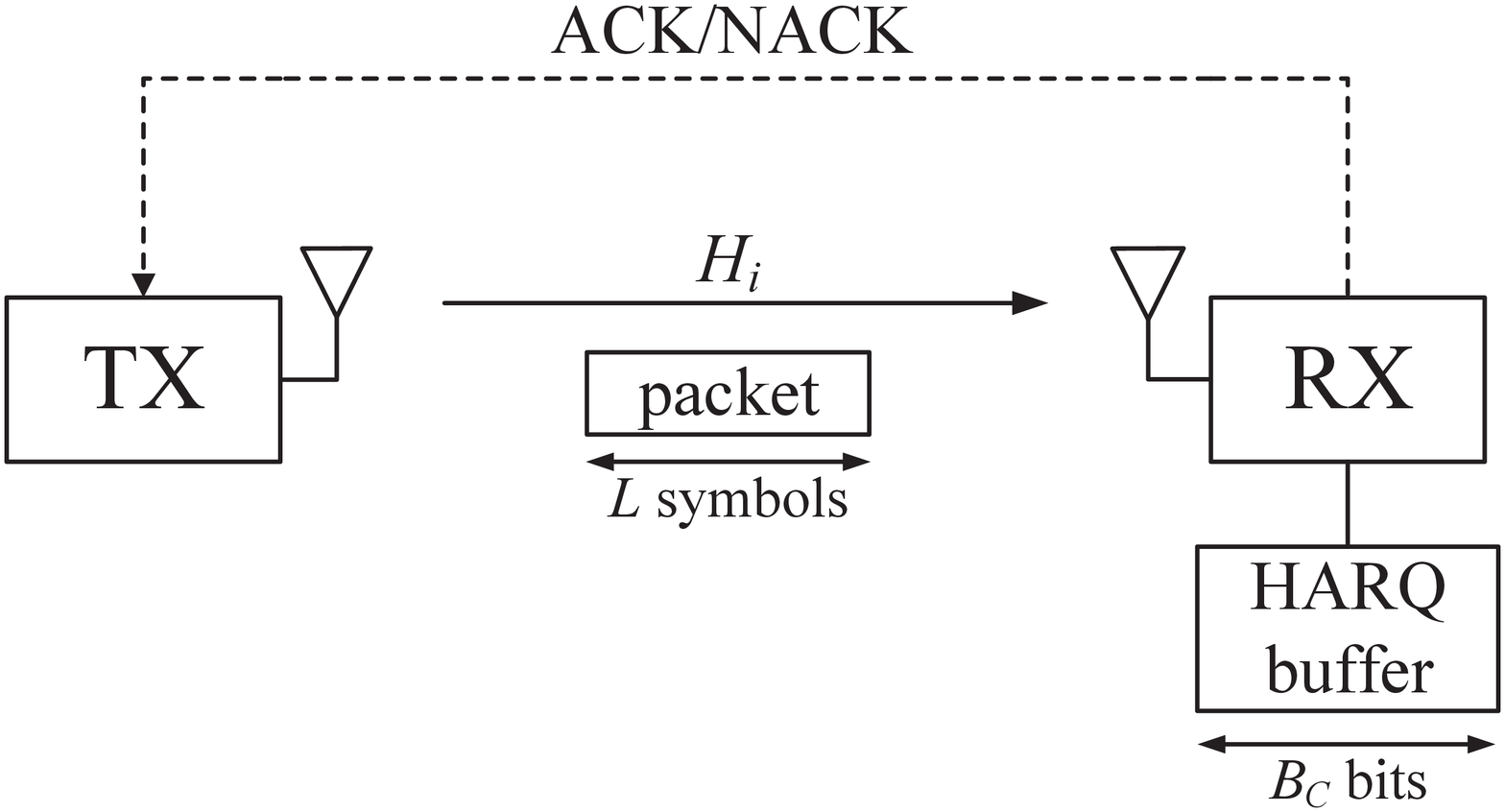, width=9cm, clip=}
\end{center}
\vspace*{-10mm} \caption{HARQ with a limited-capacity HARQ buffer. Except for Sec. \ref{ch:F_Blocklength}, we set $B_{C}=LC$, where $C$ is the buffer size normalized to the packet length.}
\label{fig:SM} 
\vspace*{-0mm}
\end{figure}

Throughout this paper, except for Sec. \ref{ch:MIMO}, we consider a communication link with a single-antenna transmitter and a single-antenna receiver operating over a quasi-static fading channel via an HARQ mechanism. As illustrated in Fig. \ref{fig:SM} and further discussed below,
we make the assumption that the receiver has a limited HARQ buffer to store information extracted from the packets received in the previous (re)transmissions.
Time is slotted and each slot accommodates the transmission of a packet of length $L$ symbols. The received signal in a channel use of the $i$-th slot is given by

\be \label{eq:system_model}
Y_{i}=\sqrt{\textrm{SNR}}H_{i}X_{i}+Z_{i},
\ee
where the parameter $\textrm{SNR}$ represents the average signal to noise ratio;
the channel gain $H_i$ has unit power and changes independently slot by slot with a given cumulative distribution function (cdf) $F$;
the input signal $X_i$ is subject to the power constraint $E[|X_{i}|^{2}]=1$; and we have the additive noise $Z_{i}\sim \mathcal{CN}(0,1)$.
The receiver has an HARQ buffer with capacity $B_{C}$ bits. Except for Sec. \ref{ch:F_Blocklength}, we will set $B_{C}=LC$, where $C$ is hence the buffer size normalized with respect to the packet length.
The channel gain $H_i$ is assumed to be known to the receiver, where, being a single (complex) value per packet, it is stored using a negligible buffer space.

Let us denote the maximum number of retransmission by $N_{max}$ and the transmission rate by $R$, which is measured in bits/s/Hz or, equivalently, in bits/symbol.
Note that, unless stated otherwise, we consider single-layer modulation at rate $R$.
The case of multi-layer modulation will be considered in Sec. \ref{ch:LayeredCoding}.
Moreover, except for Sec. \ref{ch:F_Blocklength}, the blocklength $L$ will be considered to be long enough so as to justify the use of information-theoretic asymptotic bounds.
Each HARQ session, of at most $N_{max}$ retransmission, including the original, hence aim at delivering a data packet of $LR$ bits.
For single layer modulation, the throughput $T$ can be written as (see, e.g., \cite{Steiner})
\be \label{eq:throughput}
T=\frac{R(1-P_{e}^{N_{max}})}{E[N]},
\ee
where $N$ a random variable that measures the number of retransmissions, including the original transmission, which satisfies
\be \label{eq:E_N}
E[N]=\sum_{n=1}^{N_{max}}n\textrm{Pr}[N=n];
\ee
and $P_{e}^{n}$ is the probability of an unsuccessful transmission up to, and including, the $n$-th attempt.
We have
\be
\textrm{Pr}[N=n]=P_{e}^{n-1}-P_{e}^{n}
\ee
for $n<N_{max}$ and $\textrm{Pr}[N=N_{max}]=P_{e}^{N_{max}-1}$.
Therefore, from (\ref{eq:throughput}), it is sufficient to calculate the probabilities $P_{e}^{n}$ for $n=1,\dots,N_{max}$ in order to characterize the throughput of any given HARQ scheme.
We observe that (\ref{eq:throughput}) will need to be modified to account for layered modulation.

\section{Gaussian signaling with baseband compression} \label{ch:HARQ_BBcomp}
In this section, we evaluate the throughput of HARQ-TI, HARQ-CC, and HARQ-IR
assuming a baseline scheme whereby the transmitter uses Gaussian signaling and the receiver stores in the memory compressed version of the received baseband packets.
Note that, in practice, Gaussian signaling can be interpreted as the use of an ideal coded modulation strategy at the transmitter (see, e.g., \cite{Caire}).

\subsection{HARQ-TI} \label{ch:TI_BB}
With HARQ-TI, the transmitter repeatedly sends the same encoded packet and the receiver attempts decoding based solely on the last received packet.
HARQ-TI hence does not make use of the receiver's HARQ buffer.
Under the said assumption of sufficiently large $L$, the probability of an unsuccessful transmission up to the $n$-th attempt can be obtained as
\be \label{eq:Pe_TI}
P_{e}^{n}&=&\textrm{Pr}\left[\bigcap_{i=1}^{n}\left(\textrm{SNR}|H_{i}|^2\leq 2^{R}-1\right)\right] \nonumber \\
&=&\prod_{i=1}^{n}\textrm{Pr}\left[\textrm{SNR}|H_{i}|^2\leq2^{R}-1\right] = \left(F\left(\frac{2^{R}-1}{\textrm{SNR}}\right)\right)^{n}.
\ee
We recall that the throughput is finally obtained as (\ref{eq:throughput}), which, in the case of HARQ-TI can be simplified as $T=R\left(1-P_{e}^{1}\right)$.
Note that the throughput of HARQ-TI does not depend on $N_{max}$.

\subsection{HARQ-CC} \label{ch:CC_BB}
With HARQ-CC, the transmitter repeats the same packet at each retransmission as for HARQ-TI, but the receiver performs decoding on a packet
obtained by combining all previously received packets via maximum ratio combining (MRC).
HARQ-CC hence requires storage either of all previously received packets or of the current combined packet obtained from all previous transmissions.
In the presence of a limited-buffer receiver, these two HARQ buffer management options yield different throughputs and are discussed next.

\bfig[t]
\bc
\centering \epsfig{file=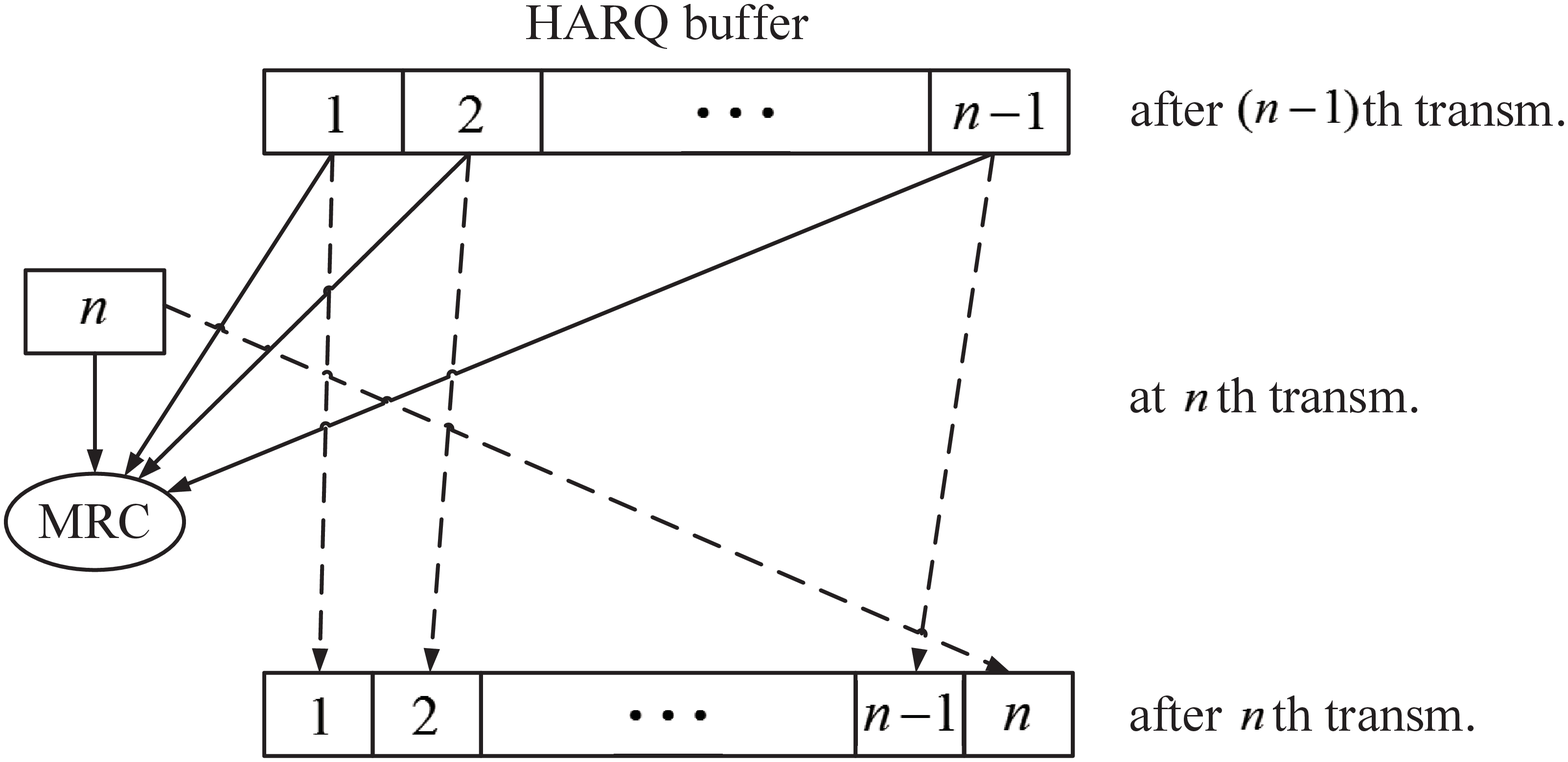, width=8cm, clip=}
\ec
\vspace*{-5mm}
\caption{Illustration of HARQ-CC S\&C. The numbers indicate the index of the packet, which is compressed in the HARQ buffer, and the dashed lines correspond to compression operations that take place in case the $n$-th transmission fails.
For the packets within the HARQ buffer recompression is carried out according to the successive refinement scheme discussed in Appendix.}
\label{fig:HARQ-CC_SC}
\efig

\subsubsection{HARQ-CC Store and Combine (S\&C)}
A first option to implement HARQ-CC in the presence of a limited HARQ buffer is for the receiver to store all the previously received packets.
Due to memory limitations, prior to storage, packets need to be compressed. To this end, as illustrated in Fig. \ref{fig:HARQ-CC_SC}, the receiver divides the available memory size equally
among all the packets received up to the given retransmissions and compresses each packet separately.
If the $n$-th transmission is unsuccessful, the receiver then compresses the last received packet to $LC/n$ bits and recompresses the previously stored packets to $LC/n$ bits (from their previous larger size of $LC/(n-1)$ bits).
We refer to this scheme as Store and Combine (S\&C).

In order to account for the effect of quantization, we use the standard additive quantization noise model.
Specifically, if the $n$-th retransmission is not successful, the quantized signals are given by
\be \label{eq:comp_packet}
\hat{Y}_{i,n}=Y_{i}+Q_{i,n},
\ee
for $i=1,\dots,n$ and $n=1,\dots,N_{max}$, where $Q_{i,n} \sim \mathcal{CN}(0,\sigma_{i,n}^{2})$ is the quantization noise for the $i$-th received packet as stored at the $n$-th unsuccessful transmission.
As discussed below, the quantization noise $\sigma_{i,n}^{2}$, which corresponds also to the mean squared error distortion, is adjusted to the current channel realization $H_i$, and hence quantization must be performed after channel estimation.

\begin{remark} \label{re:Q_model}
Quantization noise models such as (\ref{eq:comp_packet}) are used throughout this work within the information-theoretic framework of random coding, and hence the quantization noise distribution is to be considered as obtained by averaging over the randomly selected vector quantization codebooks (see, e.g., \cite{Cover,Gamal}).
Moreover, following Shannon's classical arguments, the results obtained in this paper are to be interpreted as implying the existence of specific (deterministic) coding and compression strategies that achieve the calculated throughput levels as long as they operate over sufficiently long block-lengths \cite{Cover,Gamal} (see Sec. \ref{ch:F_Blocklength} for further discussion).
From a practical viewpoint, results in \cite{Zamir} and \cite{Nagpal} suggest that high-dimensional lattice vector quantizers, such as standard Trellis Coded Quantization (TCQ) \cite{Marcellin},
or graphical codes with message passing are expected to perform close to the performance evaluated in this work.
However, the choice of a Gaussian distribution for the quantization noise is made with no claim of optimality and may be in practice justified
by the fact that dithered lattice vector quantizers are able to approximate (\ref{eq:comp_packet}) with increasing accuracy as the dimensions of the quantizer increases \cite{Zamir}.
Moreover, the Gaussian assumption implies that the performance evaluated here with baseband compression can be realized by receivers that use conventional minimum-distance decoders \cite{Lapidoth}.
\end{remark}

\bfig[t]
\bc
\centering \epsfig{file=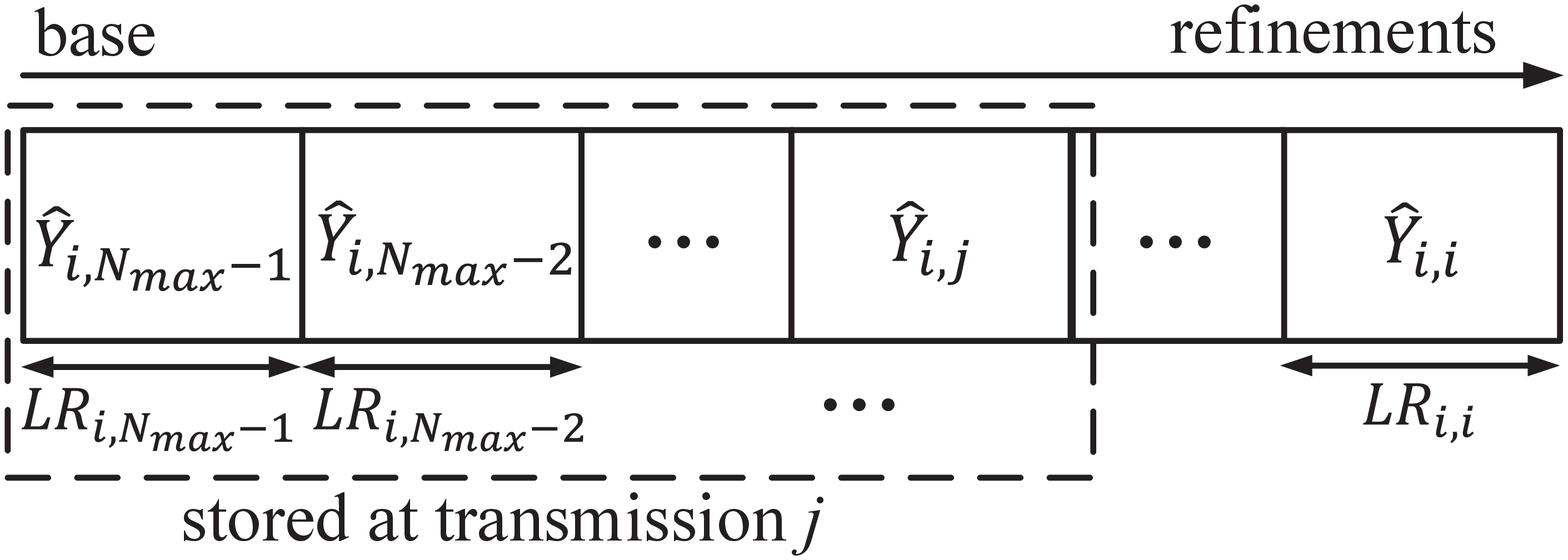, width=10cm, clip=}
\ec
\vspace*{-10mm} \caption{Illustration of the successive refinement compression strategy used for HARQ-CC S\&C and HARQ-IR: Each packet $i$ is compressed to
$(N_{max}-i)$ layers such that higher layers, corresponding to refinements, are discarded as $n$ increases in order to free memory space for the more recent packets (see, e.g. Fig. \ref{fig:HARQ-CC_SC}).}
\label{fig:App} 
\vspace*{-0mm}
\efig

Following Remark \ref{re:Q_model}, we relate the quantization noise $\sigma_{i,n}^{2}$ to the number of allocated bits $LC/n$ via the standard rate-distortion theoretic equality \cite{Cover} $C/n=I(Y_{i};\hat{Y}_{i,n})$, which can be evaluated as
\be \label{eq:Cn}
\frac{C}{n}=\log_{2}\left(1+\frac{\textrm{SNR}|H_{i}|^{2}+1}{\sigma_{i,n}^{2}}\right)
\ee
implying
\be \label{eq:rd_theory}
\sigma_{i,n}^{2} = \frac{\textrm{SNR}|H_{i}|^2+1}{2^{C/n}-1}.
\ee
The equality (\ref{eq:Cn}) holds also for recompressed packets, i.e. for all packets (\ref{eq:system_model}) with $i<n$,
as long as successive refinement compression \cite[Ch.~13]{Cover} is employed.
Specifically, as illustrated in Fig \ref{fig:App}. each packet $i$ is first compressed at the $i$-th transmission (if unsuccessful) with a number ($N_{max}-i$) of compression layers.
At later transmissions, higher layers, corresponding to refinement descriptions, are progressively discarded as $n$ increases in order to satisfy the rate constraint $C/n$ and effectively increasing the quantization noise (\ref{eq:rd_theory}).
We refer to Appendix for a detailed discussion.

At the $n$-th retransmission, the decoder performs MRC of the stored $(n-1)$ packets and of the last received packet prior to decoding as
\be \label{eq:bb_comb}
\bar{Y}_{n}=H_{n}^{*}Y_{n}+\sum_{i=1}^{n-1}H_{i}^{*}\hat{Y}_{i}.
\ee
As a result, the effective SNR is equal to
\be \label{eq:effect}
\frac{\textrm{SNR}\left(\sum_{i=1}^{n}|H_{i}|^2\right)^2}{|H_{n}|^2+\sum_{i=1}^{n-1}|H_{i}|^2\left(\sigma_{i,n}^2+1\right)},
\ee
and the probability of an unsuccessful transmission up to the $n$-th attempt is given by
\be \label{eq:Pe_QALL}
P_{e}^{n}=\textrm{Pr}\left[\bigcap_{j=1}^{n}\left(\log_{2}\left(1+\frac{\textrm{SNR}\left(\sum_{i=1}^{j}|H_{i}|^2\right)^2}{|H_{j}|^2+\sum_{i=1}^{j-1}|H_{i}|^2\left(\sigma_{i,j}^2+1\right)}\right) \leq R \right)\right].
\ee
The probability in (\ref{eq:Pe_QALL}) can be calculated via Monte Carlo simulations and the same will apply also to the other probabilities indicated in the rest of the paper.

\begin{remark} \label{re:QALL}
In the absence of memory restrictions, i.e., with $C\rightarrow \infty$, we have $P_{e}^{n}=\textrm{Pr}\left[\sum_{i=1}^{n}\textrm{SNR}|H_{i}|^2\leq2^{R}-1\right]$ since $\sigma_{i,j}^{2}\rightarrow 0$, hence obtaining the standard performance of Chase combining (see, e.g., \cite{Steiner}).
Therefore, under this conventional assumption, there is no need to include the intersection operation in (\ref{eq:Pe_QALL}).
This is because, with $C\rightarrow \infty$, the effective SNR (i.e., the ratio in (\ref{eq:Pe_QALL})) is a monotonically increasing function of $n$,
while this is generally not the case for finite $C$ due to the increasing quantization noise power (\ref{eq:rd_theory}).
\end{remark}

\begin{remark} \label{re:opDec}
The combining (\ref{eq:bb_comb}) does not account for the different noise powers affecting the combined packets due to the quantization noise.
Therefore, the combining (\ref{eq:bb_comb}) is suboptimal for finite $C$ in terms of the achievable rate.
In fact, it reflects the operation of a standard Chase combiner \cite{Chase}, which is oblivious to the presence of quantization effects.
For reference, we observe that an optimal combining would achieve an effective SNR of (see, e.g. \cite{Steiner})
\be \label{eq:op_effSNR}
\textrm{SNR}|H_{n}|^2+\sum_{i=1}^{n-1}\frac{\textrm{SNR}|H_{i}|^2}{1+\sigma_{i,n-1}^{2}},
\ee
which is generally larger than (\ref{eq:effect}) and it coincides with (\ref{eq:effect}) for $C\rightarrow \infty$.
We also refer to Sec. \ref{ch:aStore_Gau} for a discussion of an adaptive storing scheme that uses the standard Chase combiner but decides adaptively
whether to store a packet or not depending on the effective SNR improvement obtained as a result.
\end{remark}

\bfig[t]
\bc
\centering \epsfig{file=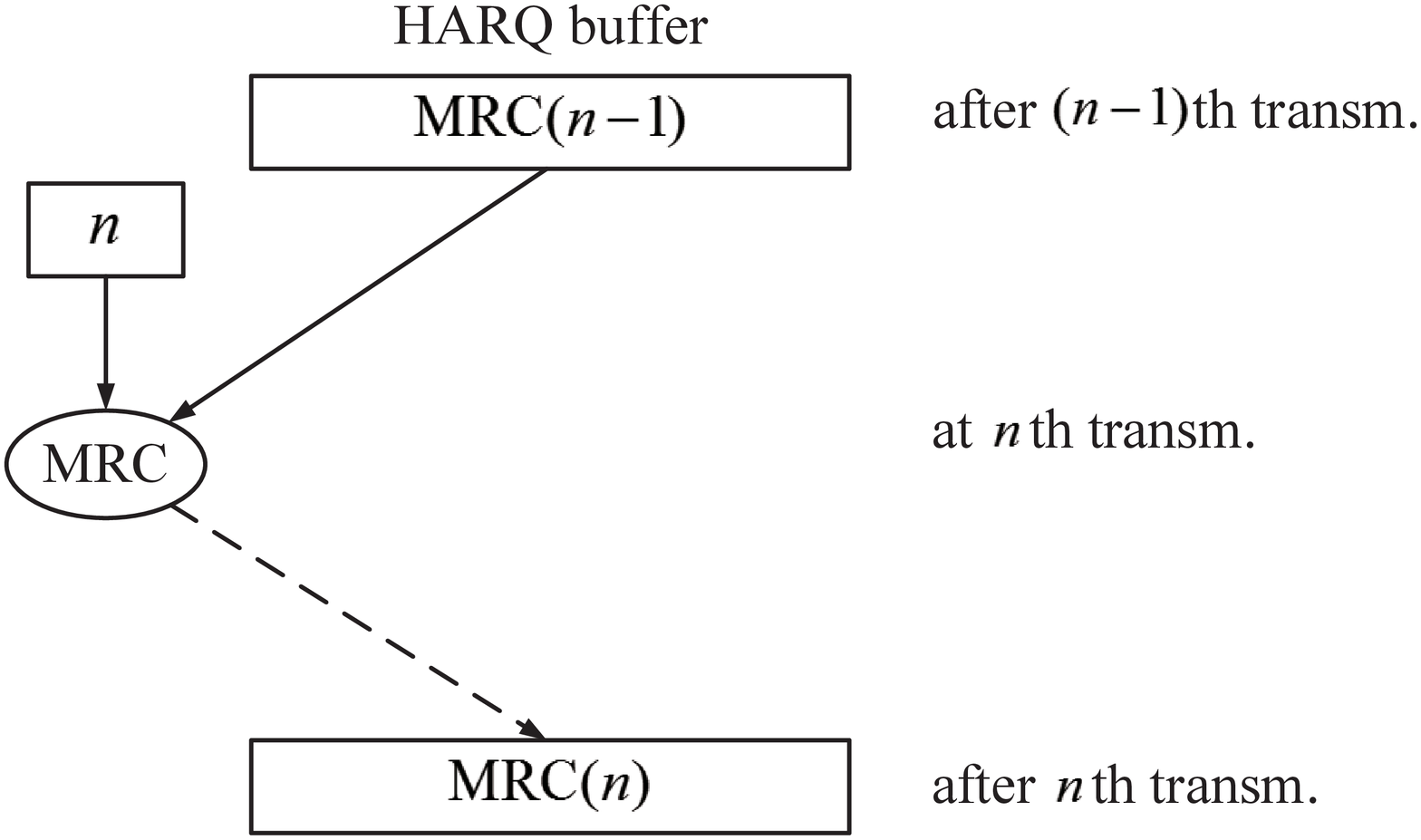, width=8cm, clip=}
\ec
\vspace*{-5mm}
\caption{Illustration of HARQ-CC C\&S. $\textrm{MRC}(n)$ indicates the compressed MRC-combined packet stored at the $n$-th transmission if unsuccessful.}
\label{fig:HARQ-CC_CS}
\efig

\subsubsection{HARQ-CC Combine and Store (C\&S)}
The S\&C approach is expected to be inefficient since decoding is carried out on the combined packet and not on the individual packets, which thus need not be separately stored.
For this reason, here, rather than storing all the previously received packets as with S\&C, we consider compressing and storing directly the MRC-combined packet.
Specifically, as illustrated in Fig. \ref{fig:HARQ-CC_CS}, at each retransmission, the last received packet is combined with the current stored packet in the HARQ buffer.
If decoding is unsuccessful, the combined packet is compressed and stored.
We refer to this scheme as Combine and Store (C\&S).

To elaborate, if decoding is not successful at the first transmission, the compressed packet is given by
\be
\hat{Y}_{1} &=& H_{1}^{*}Y_{1}+Q_{1} \nonumber \\
&=& \sqrt{\textrm{SNR}}|H_{1}|^2 X+H_{1}^{*}Z_{1}+Q_{1} \nonumber \\
&=& \sqrt{\textrm{SNR}}|H_{1}|^2 X+E_{1},
\ee
where $Q_{1}\sim\mathcal{CN}\left(0,\sigma_{1}^{2}\right)$ is the quantization noise and $E_{1}=H_{1}^{*}Z_{1}+Q_{1}\sim\mathcal{CN}\left(0,\rho_{1}^{2}\right)$ is referred to as the effective noise.
From rate-distortion theory, similar to (\ref{eq:rd_theory}), we have $\sigma_{1}^{2}=\left(|H_{1}|^{2}+\textrm{SNR}|H_{1}|^{4}\right)/\left(2^{C} -1\right)$ and $\rho_{1}^{2}=|H_{1}|^2+\sigma_{1}^{2}$.
The combined signal used in decoding at the $n$-th transmission is given by
\be \label{eq:comb_sig}
\bar{Y}_{n}=H_{n}^{*}Y_{n}+\hat{Y}_{n-1},
\ee
for all $n>1$. Moreover, the stored packet at the $n$-th attempt, if is unsuccessful, can be written as
\be \label{eq:CS}
\hat{Y}_{n} &=& \bar{Y}_{n}+Q_{n} \nonumber \\
&=& \sqrt{\textrm{SNR}}\sum_{i=1}^{n}|H_{i}|^2 X+E_{n},
\ee
with the effective noise given by $E_{n}=E_{n-1}+H_{n}^{*}Z_{n}+Q_{n}\sim \mathcal{CN}\left(0,\rho_{n}^2\right)$.
The power of the effective noise can be expressed using the recursive relationship
\be \label{eq:sigma_CQ}
\rho_{n}^{2}=\rho_{n-1}^{2}+|H_{n}|^2+\left\{\rho_{n-1}^{2}+|H_{n}|^{2}+\textrm{SNR}\left(\sum_{i=1}^{n}|H_{i}|^2\right)^{2}\right\}\bigg/\left(2^{C}-1\right).
\ee
Based on (\ref{eq:comb_sig}) and (\ref{eq:sigma_CQ}), we can finally obtain the probability of an unsuccessful transmission up to the $n$-th attempt as
\be \label{eq:Pe_CQ}
P_{e}^{n}&=&\textrm{Pr}\left[\bigcap_{j=1}^{n}\left(\log_{2}\left(1+\frac{\textrm{SNR}\left(\sum_{i=1}^{j}|H_{i}|^2\right)^2}{|H_{j}|^2+\rho_{j-1}^{2}}\right) \leq R \right)\right],
\ee
where we set $\rho_{0}=0$.

\begin{remark} \label{re:CQ}
As $C\rightarrow\infty$, the effective noise is given by $\rho_{n}^2=\sum_{i=1}^{n}|H_{i}|^2$ and we have $P_{e}^{n}=\textrm{Pr}\left[\sum_{i=1}^{n}\textrm{SNR}|H_{i}|^2\leq2^{R}-1\right]$.
The other considerations made in Remark \ref{re:QALL} and Remark \ref{re:opDec} apply here as well.
\end{remark}

\begin{remark} \label{re:MRC}
As per Remark \ref{re:opDec}, the MRC operation in (\ref{eq:CS}) neglects the fact that the noise power on the received signal at the current $n$-th transmission is different from the effective noise power $\rho_{n-1}^{2}$
that affects the previously received and combined packets due to the quantization noise.
It hence reflects the operation of a standard Chase combiner \cite{Chase}.
\end{remark}

\subsection{HARQ-IR} \label{ch:IR_BB}
With HARQ-IR, at each retransmission, the transmitter sends a packet consisting of new parity bits from a rate-compatible code and decoding is based on the concatenation of all previously received packets.
We assume that the receiver stores all the previously received packets by following the same mechanism as in HARQ-CC S\&C, and hence allocating the same buffer space to all packets.
Note that the idea of storing a combined version of the previous baseband packets as in HARQ-CC is not directly applicable to HARQ-IR.
The compressed packets at the $n$-th retransmission are given by (\ref{eq:comp_packet}) and (\ref{eq:rd_theory}).
Since with HARQ-IR the achievable rate is the sum of the achievable rates across all transmissions (see, e.g. \cite{Caire}),
the probability of an unsuccessful transmission up to the $n$-th attempt can be obtained as
\be \label{eq:Pe_IR}
P_{e}^{n}=\textrm{Pr}\left[\bigcap_{j=1}^{n} \left(\log_{2}\left(1+\textrm{SNR}|H_{j}|^{2}\right)+\sum_{i=1}^{j-1}\log_{2}\left(1+\frac{\textrm{SNR}|H_{i}|^2}{1+\left(1+\textrm{SNR}|H_{i}|^{2}\right)/\left(2^{C/(j-1)}-1\right)}\right) \leq R \right)\right].
\ee

\begin{remark} \label{re:IR}
With $C\rightarrow\infty$, we obtain $P_{e}^{n}=\textrm{Pr}\left[\sum_{i=1}^{n}\log_{2}\left(1+\textrm{SNR}|H_{i}|^2\right)\leq R\right]$ \cite{Caire} (see also Remark \ref{re:QALL} and Remark \ref{re:CQ}).
Moreover, by setting $C\rightarrow 0$ and $N_{max}=1$, the HARQ-IR throughput tends to that of HARQ-TI as it can be seen by comparing (\ref{eq:Pe_IR}) and (\ref{eq:Pe_TI}).
\end{remark}

\subsection{Adaptive Storing} \label{ch:aStore_Gau}
So far, we have assumed that all received packets are stored either individually or after MRC.
However, in order to avoid using the available HARQ buffer capacity for received packets that do not carry significant information,
one could instead store a packet only if the achievable rate is sufficiently improved as a result.
This is particulary significant since, as discussed in Remarks \ref{re:opDec}, \ref{re:CQ} and \ref{re:IR}, the rate achievable with the studied conventional HARQ schemes does not necessarily increase with the number of retransmissions $n$.
Here, we propose an \emph{adaptive} \emph{storing} strategy that is motivated by these observations.

We first describe the idea for HARQ-CC S\&C. Let us define a random variable $N_{s}(n)$ that accounts for the number of packets that have been stored prior to transmission $n+1$.
At transmission $n$, we first check if the achievable rate in the left-hand side of the inequality in (\ref{eq:Pe_QALL}) is larger than $\eta\geq1$ times the achievable rate for the previous $n-1$ transmissions, where $\eta$ is a design parameter.
If so, the packet is stored and we set $N_{s}(n+1)=N_{s}(n)+1$; if not, the packet is not stored and $N_{s}(n+1)=N_{s}(n)$.
To evaluate the performance of this scheme, in (\ref{eq:Pe_QALL}), the sums are restricted only to the indices of the $N_{s}(n)$ stored packets.
Note that $N_{s}(n)$ and the indices of the stored packets are functions of the channel gains $|H_{i}|^2$ for $i=1,\dots,n-1$.
For HARQ-CC C\&S and HARQ-IR, adaptive storing can be implemented and analyzed by following the same procedure described above, using the achievable rate appearing in the left-hand side of the inequality in (\ref{eq:Pe_CQ}) and the rate expression on the left-hand side of (\ref{eq:Pe_IR}), respectively, in lieu of (\ref{eq:Pe_QALL}).

\section{BICM with Baseband and LLR compression} \label{ch:BICM_Comp}
In this section, we consider transmission based on BICM with a fixed $M$-ary constellation $\mathcal{X}$, where $M=2^{m}$ for some integer $m$ \cite{Caire2}.
The main motivation for this investigation, beside the practical relevance of BICM, is the aim of conforming baseband compression,
as studied in the previous section, with a more conventional implementation whereby the receiver compresses the LLRs of the coded bits in the previously received packets (see, e.g., \cite{Bai}).
It is recalled that BICM maps coded bits directly on constellation points, hence facilitating the implementation and analysis of LLR processing and enabling the study of the impact of the constellation size.

Throughout this section, we make the standard assumptions of ideal interleaving, so that the $m$ bit channels can be assumed to be independent,
of a binary i.i.d. $\textrm{Ber}(1/2)$ codewords transmitted across all bit channels and of Gray mapping \cite{Caire2}.
To elaborate, we define the $j$-th bit in the binary label of $X\in\mathcal{X},~j=1,\cdots,m$, according Gray mapping, as $X(j)$, and the set
\be
\mathcal{X}_{b}^{j} =\left\{x\in X \big| X(j)=b\right\},
\ee
for $b\in \{0,1\}$, of all constellation points in which the $j$-th bit $X(j)$ equals $b$.
With these definitions and (\ref{eq:system_model}), the LLR for the $j$-th bit of a symbol within the $i$-th retransmitted packet can be written as
\be \label{eq:def_LLR}
L_{i}^{j} = \log_{2}\frac{\sum_{x\in \mathcal{X}_{1}^{j}}\exp\left(-\left|Y_{i}-\sqrt{\textrm{SNR}}H_{i}x\right|^2\right)}{\sum_{x\in \mathcal{X}_{0}^{j}}\exp\left(-\left|Y_{i}-\sqrt{\textrm{SNR}}H_{i}x\right|^2\right)}.
\ee
In the rest of this section, we first review the performance of HARQ-TI, which does not require the use of the HARQ buffer and then study the performance of HARQ-CC and HARQ-IR first with baseband compression and then with LLR compression.

\subsection{HARQ-TI}
In order to evaluate the achievable rates with BICM, we first introduce the conditional probability density function (pdf) of $Y_{i}$ given the $j$-th bit $X_{i}(j)$, which, from (\ref{eq:system_model}), is given by
\be \label{eq:f_cond_TI}
f_{Y_{i}|X_{i}(j)}(y|b)=\frac{1}{2^{m-1}}\sum_{x\in\mathcal{X}_{b}^{j}}\frac{1}{\pi}\exp(-|Y_{i}-\sqrt{\textrm{SNR}}H_{i}x|^{2}),
\ee
using the fact that all binary variables $X_{i}(j)$ are i.i.d. $\textrm{Ber}(1/2)$.
Due to joint encoding across the $m$ bit channels, an outage event takes place when the $m$ bit channels together do not support the transmission rate.
Therefore, with HARQ-TI, the probability of an unsuccessful transmission up to the $n$-th retransmission can then be calculated as (see, e.g., \cite{Rosati})
\be \label{eq:BC_Pe}
P_{e}^{n} &=& \prod_{i=1}^{n}\Pr\left[\sum_{j=1}^{m}I(X_{i}(j);Y_{i})\leq R\right] \nonumber \\
&=&\prod_{i=1}^{n}\Pr\left[\frac{1}{2}\sum_{b=0}^{1}\sum_{j=1}^{m}\int f_{Y_{i}|X_{i}(j)}(y|b)\log_{2}\frac{f_{Y_{i}|X_{i}(j)}(y|b)}{f_{Y_{i}}(y)}dy\leq R\right],
\ee
with $f_{Y_{i}}(y)=1/2\sum_{b=0}^{1}f_{Y_{i}|X_{i}(j)}(y|b)$, where the second equality follows by direct calculation of the mutual information.
While a closed-form expression for the conditional pdf $f_{Y_{i}|X_{i}(j)}(y|b)$ appears to be difficult to obtain,
this quantity, and hence also (\ref{eq:BC_Pe}), can be estimated numerically through Monte-Carlo simulations.
Note that the throughput (\ref{eq:throughput}) can be simplified as $T=R(1-P_{e}^{1})$.

\subsection{Baseband Compression}
In this subsection, we consider the performance of HARQ-CC and HARQ-IR in the presence of baseband compression.

\subsubsection{HARQ-CC} \label{ch:BICM_CC}
Similar to (\ref{eq:f_cond_TI}), in order to evaluate the performance of HARQ-CC, we first define the conditional pdf $f_{\bar{Y}_{n}|X(j)}(y|b)$
with $\bar{Y}_{n}$ being the combined packet, which is given by (\ref{eq:bb_comb}) for HARQ-CC S\&C and (\ref{eq:comb_sig}) for HARQ-CC C\&S.
In particular, for HARQ-CC S\&C, we obtain
\be \label{eq:f_cond_CC}
f_{\bar{Y}_{n}|X(j)}(y|b) = \frac{1}{2^{m-1}}\sum_{x\in\mathcal{X}_{b}^{j}}f_{\bar{\mu}_{n},\bar{\sigma}_{n}^{2}}(y),
\ee
where $f_{\bar{\mu}_{n},\bar{\sigma}_{n}^{2}}(y)=1/(\pi\bar{\sigma}_{n}^{2})\exp(-|y-\bar{\mu}_{n}|^{2}/\bar{\sigma}_{n}^{2})$ is the pdf of a complex Gaussian variable
with mean $\bar{\mu}_{n}=\sqrt{\textrm{SNR}}\sum_{i=1}^{n}|H_{i}|^{2}x$ and variance $\bar{\sigma}_{n}^2=|H_{n}|^2 +\sum_{i=1}^{n-1}\textrm{SNR}|H_{i}|^2(\sigma_{i,n-1}^{2}+1)$ using (\ref{eq:rd_theory}),
while for HARQ-CC C\&S, we have the same mean $\bar{\mu}_{n}$ and variance $\bar{\sigma}_{n}^{2}=|H_{n}|^2+\rho_{n-1}^2$ with $\rho_{n}^{2}$ in (\ref{eq:sigma_CQ}).
We can then write $P_{e}^{n}=\Pr\left[\sum_{j=1}^{m}I(X(j);\bar{Y}_{i})\leq R\right]$, which can be calculated as (\ref{eq:BC_Pe}).

\subsubsection{HARQ-IR}
Following similar arguments as for HARQ-CC and recalling Sec. \ref{ch:IR_BB}, the probability of an unsuccessful transmission up to the $n$-th attempt can be calculated as
\be \label{eq:BC_Pe_IR}
P_{e}^{n}=\Pr\left[\bigcap_{i=1}^{n}\left(\sum_{j=1}^{m}I(X_{n}(j);Y_{n})+\sum_{k=1}^{i-1}\sum_{j=1}^{m}I(X_{k}(j);\hat{Y}_{k,i})\leq R\right)\right],
\ee
where the conditional pdf $f_{Y_{i}|X_{i}(j)}$ is given by (\ref{eq:f_cond_TI}) and the conditional pdf $f_{\hat{Y}_{k,i}|X_{k}(j)}$ of the compressed packet $\hat{Y}_{k,i}$ given by $X_{k}(j)$ is equal to
$f_{\bar{\mu}_{k},\bar{\sigma}_{k,i}^{2}}(y)$ with mean $\bar{\mu}_{k}=\sqrt{\textrm{SNR}}H_{k}x$ and variance $\bar{\sigma}_{k,i}^2=\sigma_{k,i-1}^{2}+1$ in (\ref{eq:rd_theory}).

\subsection{LLR Compression}
Here we study the performance of HARQ-CC and HARQ-IR in the presence of LLR compression.

\subsubsection{HARQ-CC}
For HARQ-CC, as done in Sec. \ref{ch:BICM_CC}, we consider both compression mechanisms S\&C and C\&S.

\paragraph{HARQ-CC Store and Combine (S\&C)}
With LLR compression, similar to Sec. \ref{ch:CC_BB}, HARQ-CC S\&C divides the available memory equally to store the compressed LLRs of the previous received packets for all bits channels.
Specifically, at the $n$-th transmission, if unsuccessful, the compressed LLR for the $i$-th transmissions and bit channel $j$ is given as
\be \label{eq:LLR_SC}
\hat{L}_{i,n}^{j}=L_{i}^{j}+Q_{i,n}^{j},
\ee
for $i=1,\dots,n$ and $n=1,\dots,N_{max}$, where we follow the same standard additive quantization noise model used in Sec. \ref{ch:HARQ_BBcomp} and the quantization noise is modelled as $Q_{i,n}^{j} \sim \mathcal{N}(0,\sigma_{i,n,j}^{2})$ (see Remark \ref{re:Q_model} for a discussion on this model).
To evaluate the quantization noise variance $\sigma_{i,n,j}^{2}$, we resort to the information-theoretic equality $I(L_{i}^{j};\hat{L}_{i,n}^{j})= C/(mn)$,
which accounts for the fact that each bit channel is allocated a memory size equal to $LC/(mn)$. Since $L_{i}^{j}$ is not Gaussian, we leverage the following well-known upper bound (see, e.g. \cite[Ch.~9]{Cover})
\be \label{eq:I_upperB}
I\left(L_{i}^{j};\hat{L}_{i,n}^{j}\right) &=& I\left(L_{i}^{j};L_{i}^{j}+Q_{i,n}^{j}\right) \nonumber \\
&\leq&\frac{1}{2}\log_{2}\left(1+\frac{\textrm{var}(L_{i}^{j})}{\sigma_{i,n,j}^{2}}\right).
\ee
This bound allows us to obtain the conservative estimate of (i.e., upper bound on) the quantization noise power $\sigma_{i,n,j}^2$ by imposing the equality $1/2\log_{2}(1+\textrm{var}(L_{i}^{j})/\sigma_{i,n,j}^{2})=C/(mn)$,
which yields
\be \label{eq:Qn_SC}
\sigma_{i,n,j}^{2}=\frac{\textrm{var}(L_{i}^{j})}{\left(2^{2C/(mn)}-1\right)}.
\ee
The variance $\textrm{var}(L_{i}^{j})$ does not appear to admit a closed-form expression but it can be easily evaluated numerically.
We observe that the estimate (\ref{eq:Qn_SC}) is valid for the recompressed packets, i.e., for $i<n$, if the decoder employs successive refinement compression as discussed in Sec. \ref{ch:HARQ_BBcomp} and Appendix.

With HARQ-CC S\&C, the combined LLR for $j$-th bit at the $n$-th attempt is given by
\be \label{eq:CS_LLRcomb}
\bar{L}_{n}^{j}=L_{n}^{j}+\sum_{i=1}^{n-1}\hat{L}_{i,n}^{j},
\ee
hence summing the current LLRs with the previously compressed LLRs.
The probability of an unsuccessful transmission for HARQ-CC S\&C is finally obtained as
$P_{e}^{n}=\textrm{Pr}\left[\bigcap_{i=1}^{n}\left(\sum_{j=1}^{m}I(X_{i}(j);\bar{L}_{i}^{j})\leq R\right)\right]$, which can be written as
\be \label{eq:SCLLR_Pe}
P_{e}^{n}=\textrm{Pr}\left[\bigcap_{i=1}^{n}\left(\frac{1}{2}\sum_{j=1}^{m}\sum_{b=0}^{1}\int f_{\bar{L}_{i}^{j}|X_{i}(j)}(l|b)\log_{2}\left(\frac{f_{\bar{L}_{i}^{j}|X_{i}(j)}(l|b)}{f_{\bar{L}_{i}^{j}}(l)}\right)dl \leq R\right)\right]
\ee
and evaluated similar to (\ref{eq:BC_Pe}).

\begin{remark} \label{re:LLRcomb}
For the same reasons explained in Remark \ref{re:opDec}, the LLR combiner (\ref{eq:CS_LLRcomb}) is optimal only when there are no HARQ buffer size limitations
and it reflects the performance of a standard combiner.
\end{remark}

\paragraph{HARQ-CC Combine and Store (C\&S)}
Instead of storing all the previously received LLRs, similar to Sec. \ref{ch:CC_BB}, HARQ-CC C\&S stores the compressed value of the combined LLRs at each transmission.
Specifically, if decoding of the first transmission is not successful, the stored LLR is given by
\be
\hat{L}_{1}^{j} = L_{1}^{j}+Q_{1}^{j},
\ee
where $Q_{1}^{j}\sim\mathcal{N}\left(0,\sigma_{1,j}^{2}\right)$ is the quantization noise.
From the information-theoretic upper bound used in (\ref{eq:I_upperB}), we have $\sigma_{1,j}^{2}=\textrm{var}(L_{1}^{j})/\left(2^{2C/m}-1\right)$.
Similar to (\ref{eq:CS_LLRcomb}), combined LLR at the $n$-th attempt can be written as
\be \label{eq:LLR_CS}
\bar{L}_{n}^{j} &=& L_{n}^{j}+\hat{L}_{n-1}^{j}
\ee
for all $m>1$, which corresponds to the optimal combiner in the absence of quantization noise (see Remark \ref{re:LLRcomb}).
Moreover, if the $n$-th attempt is unsuccessful, the compressed combined LLR is given as $\hat{L}_{n}^{j}=\bar{L}_{n}^{j}+Q_{n}^{j}$,
where $Q_{n}^{j} \sim \mathcal{N}(0,\sigma_{n,j}^2)$ with quantization noise power $\sigma_{n,j}^{2}=\textrm{var}(\bar{L}_{n}^{j})/\left(2^{2C/m}-1\right)$, since HARQ-CC C\&S allocates the available memory to store only the currently combined LLR (\ref{eq:LLR_CS}).
Similar to (\ref{eq:SCLLR_Pe}), the probability of an unsuccessful transmission up to the $n$-th retransmission is finally obtained as
$P_{e}^{n}=\textrm{Pr}\left[\bigcap_{i=1}^{n}\left(\sum_{j=1}^{m}I(X_{i}(j);\bar{L}_{i}^{j})\leq R\right)\right]$, which yields
\be
P_{e}^{n}=\textrm{Pr}\left[\bigcap_{i=1}^{n}\left(\frac{1}{2}\sum_{j=1}^{m}\sum_{b=0}^{1}\int f_{\bar{L}_{i}^{j}|X_{i}(j)}(l|b)\log_{2}\left(\frac{f_{\bar{L}_{i}^{j}|X_{i}(j)}(l|b)}{f_{\bar{L}_{i}^{j}}(l)}\right)dl \leq R\right)\right].
\ee

\subsubsection{HARQ-IR}
With HARQ-IR, as discussed in Sec. \ref{ch:IR_BB}, the transmitter sends new parity bits at each transmission and
the receiver stores the previously received LLRs by allocating the available memory as done for HARQ-CC S\&C.
Therefore, the compressed LLRs are given as (\ref{eq:LLR_SC}) with (\ref{eq:Qn_SC}).
Moreover, using the fact that the achievable rate is the sum of all achievable rates in previously received packets \cite{Caire},
the probability of an unsuccessful transmission up to the $n$-th attempt can be calculated as
$P_{e}^{n}=\textrm{Pr}\left[\bigcap_{i=1}^{n}\left(\sum_{j=1}^{m}\left(I(X_{i}(j);L_{i,i}^{j})+\sum_{k=1}^{i-1}I(X_{k}(j);\hat{L}_{k,i}^{j})\right)\leq R\right)\right]$, which yields
\be \label{eq:LLR_Pe_IR}
P_{e}^{n}&=&\textrm{Pr}\left[\bigcap_{i=1}^{n}\left(\frac{1}{2}\sum_{j=1}^{m}\sum_{b=0}^{1}\int f_{L_{i,i}^{j}|X_{i}(j)}(l|b)\log_{2}\left(\frac{f_{L_{i,i}^{j}|X_{i}(j)}(l|b)}{f_{L_{i,i}^{j}}(l)}\right)dl \right. \right. \nonumber \\
&&~~~~~~~~+ \left.\left.\frac{1}{2}\sum_{k=1}^{i-1}\sum_{j=1}^{m}\sum_{b=0}^{1}\int f_{\hat{L}_{k,i}^{j}|X_{k}(j)}(l|b)\log_{2}\left(\frac{f_{\hat{L}_{k,i}^{j}|X_{k}(j)}(l|b)}{f_{\hat{L}_{k,i}^{j}}(l)}\right)dl \leq R \right) \right].
\ee


\section{Layered Coding} \label{ch:LayeredCoding}
So far, we have made the standard assumption that each HARQ session aims at transmitting a single data packet (carrying $LR$ bits).
Here, instead, we investigate the potential throughput gains that can be achieved via layered coding \cite{Steiner}.
With layered coding, the transmitter encodes multiple data packets, each with a different data rate, using separate codebooks.
The encoded layers are superimposed to yield the transmitted signal.
Depending on the channel conditions, by the end of the HARQ session, the receiver may be able to decode only a subset of the layers.
Specifically, the receiver attempts decoding starting from the first layer up to the last using a successive cancellation procedure in which higher layers are treated as noise when decoding lower layers.
Layered coding is typically used to encode multimedia information sources compressed using successive refinement techniques, whereby the lower layers encode the most significant source description (see e.g., \cite{Ng}).
Moreover, multi-layer transmission appears to be particularly well suited to system with HARQ buffer size limitations since decoded layers can be transferred off chip and need not be retransmitted.

To elaborate, if the information rate of layer $i$ is $R_{i}$ and there are $N_{L}$ layers,
the throughput $T$ can be written as
\be \label{eq:T_LC}
T=\sum_{i=1}^{N_{L}}\frac{R_{i}(1-P_{e,i}^{N_{max}})}{E[N]},
\ee
where $P_{e,i}^{n}$ is the probability that layer $i$ has not been successfully decoded up to, and including, the $n$-th retransmission;
the number of retransmissions $N$ is given by (\ref{eq:E_N}) with
\be
\textrm{Pr}[N=n]=P_{e,N_{L}}^{n-1}-P_{e,N_{L}}^{n}
\ee
for $n<N_{max}$ and $\textrm{Pr}[N=N_{max}]=P_{e,N_{L}}^{N_{max}-1}$.
Note that the throughput (\ref{eq:T_LC}) counts as useful any successfully decoded layer of information irrespective of whether, by the end of HARQ session, all the $N_{L}$ layers are correctly decoded.

In the rest of this section, we study the throughput achievable with layered coding focusing, for simplicity of notation, on Gaussian signaling with baseband compression.
The extension to BICM can be carried out by following the same considerations as in the previous section.
Moreover, we limit the presentation to HARQ-TI and HARQ-IR. The analysis for HARQ-CC can be also performed following similar steps.
Finally, similar to \cite{Steiner}, we assume $N_{L}=2$ layers\footnote{However, unlike \cite{Steiner}, we allow for an arbitrary number $N_{max}$ of transmissions.},
but the generalization to any number of layers is straightforward albeit cumbersome in terms of notation.

\subsection{Throughput Analysis}
In order to evaluate the throughput, let us define as $K$ the random variable indicating the transmission at which the first layer is decoded correctly.
Note that we have $1\leq K \leq N_{max}$. In the following, we consider HARQ-IR and observe that the performance with HARQ-TI can be obtained by setting $C \rightarrow 0$ and $N_{max}=1$ (see Remark \ref{re:IR}).

We first fix $K=k\in\{1,\dots,N_{max}\}$ and develop the expressions for the relevant signals for the given value $K=k$.
With $N_{L}=2$ layers, the signal transmitted at the $n$-th transmission is given by the superposition
\be
X_{n,k} = \left\{
\begin{array}{lr}
\sqrt{\alpha \textrm{SNR}}X_{n}^{(1)}+\sqrt{(1-\alpha) \textrm{SNR}}X_{n}^{(2)} & \textrm{if} ~ n \leq k, \\
\sqrt{\textrm{SNR}}X_{n}^{(2)}& \textrm{if} ~ n > k,
\end{array}
\right.
\ee
where $X_{n}^{(i)} \sim \mathcal{CN}(0,1)$ is the signal encoding layer $i$ at the $n$-th transmission and $\alpha$ is a power splitting factor with $0 \leq \alpha \leq 1$.
All signals $X_{n}^{(i)}$ for $i=1,2$ and $n=1,\dots,N_{max}$ are independent.
Note that we have made the dependence of the transmitted signal on $K=k$ explicit.
The received signal at the $n$-th transmission is given, if $K=k$, by $Y_{n,k}=H_{n}X_{n,k}+Z_{n}$, where $Z_{n}\sim \mathcal{CN}(0,1)$.

With HARQ-IR, at the $n$-th retransmission, the previously received packets are compressed and stored by allocating the available memory equally across all packets as in Sec. \ref{ch:HARQ_BBcomp} and Sec. \ref{ch:BICM_Comp}.
As a result, the compressed $i$-th packet at the transmission $n\geq i$, if the first layer is decoded at the $k$-th transmission, is given by
\be
\hat{Y}_{i,n,k}=Y_{i,k}+Q_{i,n,k},
\ee
for $i=1,\dots,n-1$, where $Q_{i,n,k}\sim\mathcal{CN}(0,\sigma_{i,n,k}^{2})$ is the additive quantization noise.
Similar to (\ref{eq:rd_theory}), from rate-distortion theory, the variance $\sigma_{i,n,k}^{2}$ can be obtained as
\be \label{eq:LC_IR}
\sigma_{i,n,k}^{2}=\left\{
\begin{array}{lr}
\left(\textrm{SNR}|H_{i}|^2+1\right)\big/\left(2^{C/n}-1\right)& \textrm{if} ~ n \neq k, \\
\left((1-\alpha)\textrm{SNR}|H_{i}|^2+1\right)\big/\left(2^{C/n}-1\right)& \textrm{if} ~ n = k. \\
\end{array}
\right.
\ee
Note that, for $n=k$, the first layer is removed prior to compression and hence the power of the signal to be compressed is reduced.
Moreover, we observe that cancellation of the first layer does not decrease the quantization noise of the packets that have been already compressed.

The sum of the achievable rates, i.e., mutual informations, for the first layer at the transmission $n\leq k$, can be obtained as
\be \label{eq:LC_IR_R1}
R_{1}(n)&=&\log_{2}\left(1+\frac{\alpha \textrm{SNR}|H_{n}|^2}{1+(1-\alpha)\textrm{SNR}|H_{n}|^2}\right)+\sum_{i=1}^{n-1}\log_{2}\left(1+\frac{\alpha \textrm{SNR}|H_{i}|^2}{1+\sigma_{i,n-1,k}^2+(1-\alpha)\textrm{SNR}|H_{i}|^2}\right),
\ee
in which the second layer is treated as additional noise, along with the quantization nose.
Note that the rate $R_{1}(n)$ in (\ref{eq:LC_IR_R1}) is statistically independent of $K$ due to the definition (\ref{eq:LC_IR})
and hence we have only emphasized the dependence on $n$.
Similarly, the accumulated rate for the second layer at the $n$-th retransmission for $n \geq k$ can be written as (see also \cite{Steiner})
\be \label{eq:LC_IR_R2}
R_{2}(n,k) =\left\{
\begin{array}{lr}
\log_{2}\left(1+(1-\alpha)\textrm{SNR}|H_{n}|^2\right)+\sum_{i=1}^{n-1}\log_{2}\left(1+\frac{(1-\alpha)\textrm{SNR}|H_{i}|^2}{1+\sigma_{i,n-1,k}^2}\right)& \textrm{if} ~ n = k, \\
\log_{2}\left(1+\textrm{SNR}|H_{n}|^2\right)+\sum_{i=1}^{k}\log_{2}\left(1+\frac{(1-\alpha)\textrm{SNR}|H_{i}|^2}{1+\sigma_{i,n-1,k}^2}\right) \\ \hspace{50mm}+\sum_{i=k+1}^{n-1}\log_{2}\left(1+\frac{\textrm{SNR}|H_{i}|^2}{1+\sigma_{i,n-1,k}^2}\right) & \textrm{if} ~ n > k.
\end{array}
\right.
\ee
Note that (\ref{eq:LC_IR_R2}) accounts for the facts that the second layer is considered for decoding only after the first layer is decoded,
and that the first layer is cancelled from the received signal prior to decoding of the second layer.
We also remark that $R_{2}(n,k)$ depends on the value $K=k$.

The probability of an unsuccessful transmission for the first layer at the $n$-th transmission is given by
\be
P_{e,1}^{n} = \Pr\left[\bigcap_{i=1}^{n} \left(R_{1}(i)<R_{1}\right)\right],
\ee
where the probability is taken, here and for the rest of this section, with respect to the distribution of the channel discussed in Sec. \ref{ch:systemmodel}.
The probability of an unsuccessful transmission for the second layer at the $n$-th transmission is given by
\be
P_{e,2}^{n} &=&\sum_{k=1}^{n}\Pr\left[K=k\right]\Pr\left[\bigcap_{j=k}^{n}\left(R_{2}(j,k)<R_{2}\right)\bigg|K=k\right]\nonumber \\
&=&\sum_{k=1}^{n}\Pr\left[K=k\right]\prod_{j=k}^{n}q(j,k)\nonumber \\
&=&\sum_{k=1}^{n}\left(P_{e,1}^{k-1}-P_{e,1}^{k}\right)\prod_{j=k}^{n}q(j,k),
\ee
where the first equation follows from the law of total probability and the second from the chain rule with the definition
\be
q(j,k)=\Pr\left[R_{2}(j,k)<R_{2}\Bigg| \bigcap_{i=1}^{k-1}\left(R_{1}(i)<R_{1}\right)\bigcap \left(R_{1}(k)\geq R_{1}\right) \bigcap_{i=k}^{j-1} \left(R_{2}(i,k)<R_{2}\right)\right].
\ee

\section{Multiple-antenna links} \label{ch:MIMO}
While the analysis has focused so far on single-antenna systems, in this section we elaborate on some of the additional challenges and opportunities
that arise in the design of compression for HARQ buffer management when considering multiple-antenna, or MIMO, links.
Specifically, we consider a MIMO link with $N_{t}$ transmit antennas and $N_{r}$ receive antennas.
The $N_{r} \times 1$ received vector at each symbol of the $i$-th retransmission can be written as
\be
\mathbf{Y}_{i} = \sqrt{\textrm{SNR}}\mathbf{H}_{i}\mathbf{X}_{i}+\mathbf{Z}_{i},
\ee
where $\textrm{SNR}$ is the average signal to noise ratio per receive antenna;
the $N_{r}\times N_{t}$ channel matrix $\mathbf{H}_{i}$ has unit power elements and changes independently at each retransmission;
the $N_{t} \times 1$ vector of transmitted symbols $\mathbf{X}_{i}$ has unit average power, i.e., $E[\|\mathbf{X}_{i}\|^2]=1$;
and we have the additive noise $\mathbf{Z}_{i}\sim\mathcal{CN}(0,\mathbf{I})$.
We focus on Gaussian signaling, by setting $\mathbf{X}_{i} \sim \mathcal{CN}(0,\mathbf{I}/N_{t})$, on baseband compression and, for its relevance, on HARQ-IR.
We also assume single-layer transmission and sufficiently large blocklengths so as to invoke standard information theoretic results.
Extensions are left for future work.

As done throughput this paper, we assume that the receiver compresses and stores the packets by equally dividing the HARQ buffer.
However, while in the single-antenna case, under the additive Gaussian quantization noise model, this allocation fully determines the quantization noise power,
and hence the quantization strategy, with a multiple-antenna receiver a new design degree of freedom arises.
Specifically, the designer can control the correlation of the additive Gaussian quantization noise across the received antennas.
As discussed in, e.g., \cite{Zamir,Del_coso}, such correlation can be equivalently realized via a transform coding strategy, whereby the received signal is first processed by a linear transform and then independent noise is added to the elements of the resulting signal.
We elaborate on this approach and on the optimization of the linear transform in the rest of this section.

If the $n$-th retransmission is not successful, the signal $\mathbf{Y}_{i}$ received at the $i$-th transmission is compressed$-$for the first time if $i=n$ or recompressed, by removing the current enhancement layer (Fig. \ref{fig:App}), if $i<n$$-$ as
\be \label{eq:hat_y}
\hat{\mathbf{Y}}_{i,n} = \mathbf{A}_{i,n}\mathbf{Y}_{i}+\mathbf{Q}_{i},
\ee
where $\textbf{A}_{i,n}$ is a transform coding matrix to be calculated
and $\mathbf{Q}_{i}\sim\mathcal{CN}(0, \mathbf{I})$ is the vector of independent Gaussian quantization noises.
Note that model (\ref{eq:hat_y}) is consistent with the assumed successive refinement strategy (see Fig. \ref{fig:App})
only if the transforms $\mathbf{A}_{i,n}$ are selected so that the Markov chain $\mathbf{Y}_{i}-\hat{\mathbf{Y}}_{i,i}-\hat{\mathbf{Y}}_{i,i+1}\cdots-\hat{\mathbf{Y}}_{i,N_{max}-1}$ is preserved (see Appendix).
This will be ensured by the strategy proposed below.

Assuming joint encoding across all transmission antennas (see, e.g., \cite{Lozano}), the achievable rate of HARQ-IR, at the $n$-th attempt, can be written as the sum of the mutual informations
\be \label{eq:ach_R}
I(\mathbf{X}_{n};\mathbf{Y}_{n})+\sum_{i=1}^{n-1}I(\mathbf{X}_{i};\hat{\mathbf{Y}}_{i,n}).
\ee
Therefore, we propose to design the transform matrix $\mathbf{A}_{i,n}$ so as to optimize (\ref{eq:ach_R}) under the constraint that the HARQ buffer is equally allocated to all packets.
Defining $\mathbf{\Omega}_{i,n}=\mathbf{A}_{i,n}^{\dagger}\mathbf{A}_{i,n}$, the optimization problem is stated as
\be
\textrm{maximize}_{\mathbf{\Omega}_{i,n}\succeq 0}~I(\mathbf{X}_{i};\hat{\mathbf{Y}}_{i,n})=\log\det \left(\mathbf{I}+\mathbf{\Omega}_{i,n}\left(\mathbf{I}+\frac{\textrm{SNR}}{N_{t}}\mathbf{H}_{i}\mathbf{H}_{i}^{\dagger}\right)\right)-\log\det\left(\mathbf{I}+\mathbf{\Omega}_{i,n}\right)\nonumber \\
\textrm{s.t.}~ I(\mathbf{Y}_{i};\hat{\mathbf{Y}}_{i,n})=\log\det \left(\mathbf{I}+\mathbf{\Omega}_{i,n}\left(\mathbf{I}+\frac{\textrm{SNR}}{N_{t}}\mathbf{H}_{i}\mathbf{H}_{i}^{\dagger}\right)\right) \leq \frac{C}{n-1}. \hspace{14mm}
\ee
Following \cite{Del_coso}, given the eigenvalue decomposition \sloppy{$\mathbf{I}+(\textrm{SNR}/N_{t})\mathbf{H}_{i}\mathbf{H}_{i}^{\dagger} = \mathbf{U}\textrm{diag}\left(\lambda_{i,1},\dots,\lambda_{i,N_{r}}\right)\mathbf{U}^{\dagger}$ with unitary matrix $\mathbf{U}$} and
ordered eigenvalues $\lambda_{i,1}\geq\cdots\geq\lambda_{i,N_{r}}$, an optimal solution is given by $\mathbf{\Omega}_{i,n}^{*}= \mathbf{U}\textrm{diag}\left(\alpha_{i,n,1},\dots,\alpha_{i,n,N_{r}}\right)\mathbf{U}^{\dagger}$ with
\be \label{eq:op_alpha}
\alpha_{i,n,l}=\left[\frac{1}{\mu_{n}}\left(1-\frac{1}{\lambda_{i,l}}\right)-1\right]^{+},
\ee
where Lagrangian multiplier $\mu_{n}$ is selected so that the condition
\be \label{eq:const_buffer}
\sum_{l=1}^{N_{r}}\log(1+\alpha_{i,n,l}\lambda_{i,l})=\frac{C}{n-1}
\ee
is satisfied. We observe that (\ref{eq:op_alpha})-(\ref{eq:const_buffer}) guarantee that the gains $\alpha_{i,n,l}$ for $l=1,\dots,N_{r}$ are non-increasing functions of the right-hand side of (\ref{eq:const_buffer}).
This can be seen to imply the Markov chain mentioned above and hence the feasibility of successive refinement, as further elaborated in the Remark below.

\begin{remark}
The transform coding compression strategy (\ref{eq:hat_y}) under the proposed optimal design prescribes the choice of matrix $\mathbf{A}_{i,n}$ as
\be
\mathbf{A}_{i,n}=\textrm{diag}\left(\sqrt{\alpha_{i,n,1}},\dots,\sqrt{\alpha_{i,n,N_{r}}}\right)\mathbf{U}_{i}^{\dagger}.
\ee
This can be in practice accomplished by multiplying the received signal by the orthogonal transform matrix $\mathbf{U}_{i}^{\dagger}$ and then
multiplying the entries of the resulting vector by the corresponding gains $\sqrt{\alpha_{i,n,l}}$ prior to compression with independent unit-power quantization noises.
Note that the matrix $\mathbf{U}_{i}$ is the Karhunen-Loeve transform for the received signal and hence the output vector has independent entries that can be independently quantized with no loss of optimality (see e.g., \cite{Del_coso,Park}).
We also observe that the fact that the gain $\alpha_{i,n,l}$ is non-increasing with respect to $n=i,\dots,N_{max}-1$ for every $l=1,\dots,N_{r}$ proves that
the Markov chain $\mathbf{Y}_{i}-\hat{\mathbf{Y}}_{i,i}-\hat{\mathbf{Y}}_{i,i+1}\cdots-\hat{\mathbf{Y}}_{i,N_{max}-1}$ holds and hence successive refinement can be employed as discussed in Sec. \ref{ch:HARQ_BBcomp} and detailed in the Appendix.
\end{remark}

With the optimal solution $\mathbf{\Omega}_{1,n}^{*},\dots,\mathbf{\Omega}_{n-1,n}^{*}$ based on (\ref{eq:op_alpha}),
the probability of an unsuccessful transmission up to the $n$-th retransmission is obtained as
\be
P_{e}^{n} &=& \Pr\left[I(\mathbf{X}_{n};\mathbf{Y}_{n})+\sum_{i=1}^{n-1}I(\mathbf{X}_{i};\hat{\mathbf{Y}}_{i,n})<R \right] \nonumber \\
&=& \Pr\Bigg[\log\det \left(\mathbf{I}+\frac{\textrm{SNR}}{N_{t}}\mathbf{H}_{i}\mathbf{H}_{i}^{\dagger}\right) +\sum_{i=1}^{n-1}\log\det \left(\mathbf{I}+\mathbf{\Omega}_{i,n}^{*}\left(\mathbf{I}+\frac{\textrm{SNR}}{N_{t}}\mathbf{H}_{i}\mathbf{H}_{i}^{\dagger}\right)\right)\nonumber \\
&&\hspace{60mm}-\log\det\left(\mathbf{I}+\mathbf{\Omega}_{i,n}^{*}\right)<R \Bigg],
\ee
which can be used in (\ref{eq:throughput}) to evaluate the throughput.

\section{Optimizing the Blocklength} \label{ch:F_Blocklength}
In the previous sections, we made the classical assumption that the blocklength $L$ is large enough so as to be able to invoke
the asymptotic information-theoretic characterizations for achievable communication and compression rates.
In this section, we instead turn to the investigation of the impact of the selection of the blocklength $L$ on the HARQ throughput.
This study is motivated by the facts that a large $L$ generally entails a smaller probability of error and a more effective (vector) quantization but it also requires the storage of more information in the HARQ buffer.
An optimal value of $L$ is hence expected to result from the trade-off between these effects.

In order to study the impact of the blocklength $L$, we leverage recent \sloppy{information-theoretic} studies on the finite-blocklength performance of channel coding \cite{Polyanskiy} and source coding \cite{Kostina}.
In this section, our approach is based on the same type of approximation proposed in \cite{Devassy} that are motivated by the studies \cite{Polyanskiy}, \cite{Yang}.
Furthermore, to account for the possibility to optimize the blocklength $L$, we consider the size of the HARQ buffer to be described by the total number of bits $B_{C}$ that it can store (and not by the normalized value $C$).
In this fashion, an increase in $L$ does not entail a larger HARQ buffer.
We define as $b$ the total number of bits to be communicated in an HARQ session.
Finally, similar to the previous section, we focus on the performance of HARQ-IR for a single-antenna link with Gaussian signaling and baseband compression,
with the understanding that setting $B_{C}\rightarrow 0$ and $N_{max}=1$ yields the performance of HARQ-TI.

\subsection{Throughput Analysis}
As done throughout this paper, for HARQ-IR, we assume that the receiver compresses and stores every packet by allocating an equal fraction of the HARQ buffer to all stored packets.
In order to account for the effect of a finite blocklength $L$ on the performance of the compressor, we leverage the main results in \cite{Kostina}.
Accordingly, for a given tolerated performance $\epsilon_{q}$ that an optimal quantizer fails to compress a Gaussian signal with power $P$ and a given quantization noise variance $\sigma^2$,
the necessary storage space is approximately given by \cite{Kostina}
\be \label{eq:rd_FBL}
L\left(\log_{2}\left(1+\frac{P}{\sigma^{2}}\right)+\sqrt{\frac{V_{q}}{L}}Q^{-1}(\epsilon_{q})\right),
\ee
where the rate-dispersion factor $V_{q}$ is defined as $V_{q}=1/2\log_{2}^{2}e$.
Note that the term $\sqrt{V_{q}/L}Q^{-1}(\epsilon_{q})$ measures the redundancy due to finite blocklength effects
and that this redundancy increases with a smaller probability of compression error $\epsilon_{q}$.
Moreover, we observe that a quantizer failure can be detected by calculating the resulting distortion.

In order to apply the result (\ref{eq:rd_FBL}) to the analysis of HARQ-IR, we observe the following.
First, the number $N_{s}(n)$ of successfully compressed, and hence stored, packets prior to the $(n+1)$-th transmission is a random variable whose distribution depends on the selection of $\epsilon_{q}$.
Here, we assume that each packet $Y_{i}$ at transmission $i$, in case of unsuccessful decoding, is stored with probability $1-\epsilon_{q}$ or discarded due to a compression failure with probability $\epsilon_{q}$, independently for all $i=1,\dots,N_{max}$.
The independence assumption follows from the independence of the signals $Y_{i}$, $i=1,\dots,N_{max}$.
As a result, the variable $N_{s}(n)$ is binomial with parameter $n$ and $1-\epsilon_{q}$.
A second comment is that (\ref{eq:rd_FBL}) applies also in the presence of successive refinement
since, with Gaussian sources and mean squared error distribution, successive refinement can be optimally performed by quantizing at each layer the residual between the source and the previous coarser description, which is also a Gaussian source \cite{Gamal}.

For a given number $N_{s}(n)$ of previously stored packets, if $L$ is large enough, the transmission rate of HARQ-IR at the $n$-th transmission can be written as
\be \label{eq:R_IR_FBL}
R(n) &=& \log_{2}\left(1+\textrm{SNR}|H_{n}|^{2}\right)+\sum_{i=1}^{N_{s}(n-1)}\log_{2}\left(1+\frac{\textrm{SNR}|H_{i}|^2}{1+\sigma_{i,N_{s}(n-1)}^{2}}\right),
\ee
where $\sigma_{i,N_{s}(n-1)}^{2}$ is obtained from (\ref{eq:rd_FBL}) as
\be \label{eq:sigma_FBL}
\sigma_{i,N_{s}(n-1)}^{2} = \frac{\textrm{SNR}|H_{i}|^2 +1}{2^{B_{C}/(LN_{s}(n-1))-\sqrt{V_{q}/L}Q^{-1}(\epsilon_{q})}-1}.
\ee

Thus, by using (\ref{eq:R_IR_FBL}) and the Gaussian approximation used in \cite{Devassy}, inspired by \cite{Polyanskiy}, \cite{Yang},
the probability of an unsuccessful transmission up to the $n$-th attempt can be approximated as
\be \label{eq:Pe_IR_FBL}
P_{e}^{n} \approx E\left[Q\left(\frac{R(n)+1/(2L)\log_{2}L-b/L}{\sqrt{V_{c}/L}}\right)\right],
\ee
where $b$ is the number of information bits; the channel dispersion function $V_{c}$ is defined as
\be
V_{c} = \left(N_{s}(n-1)+1-\left(1+\textrm{SNR}|H_{n}|^{2}\right)^{-2}-\sum_{i=1}^{N_{s}(n-1)}\left(1+\frac{\textrm{SNR}|H_{i}|^2}{1+\sigma_{i,N_{s}(n-1)}^{2}}\right)^{-2}\right)\log_{2}^2e;
\ee
and the expectation is taken over the channel and the variable $N_{s}(n-1)$, which are mutually independent.
Using (\ref{eq:Pe_IR_FBL}) in (\ref{eq:throughput}) yields an approximation on the achievable throughput, which will be taken here as the performance metric of interest.

\section{Numerical Results} \label{ch:simulation}
In this section, we evaluate the throughput performance of HARQ in the presence of a finite buffer under Rayleigh fading, i.e., all channels $H_{i}$ are independent zero-mean unit-power complex Gaussian variables, via numerical results.
We first assume standard single-layer transmission and a large blocklength as studied in Sec. \ref{ch:HARQ_BBcomp} and Sec. \ref{ch:BICM_Comp},
and then we consider the impact of layered coding and of an optimized blocklength as investigated in Sec. \ref{ch:LayeredCoding} and Sec. \ref{ch:F_Blocklength}, respectively.
Lastly, we study the optimal quantization strategy for a MIMO link as proposed in Sec. \ref{ch:MIMO}.

\bfig[t]
\bc
\centering \epsfig{file=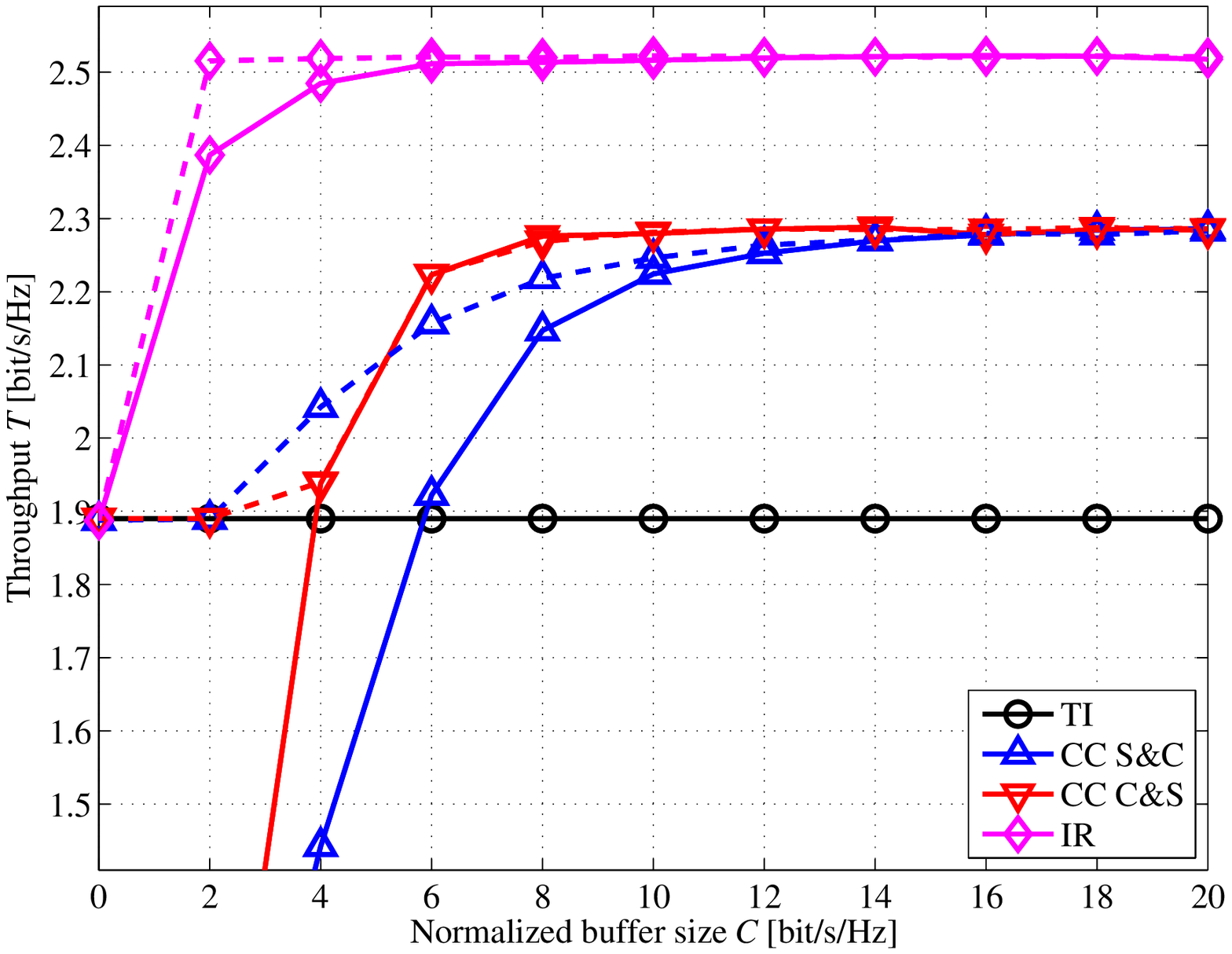, width=13cm, clip=}
\ec
\vspace*{-10mm}
\caption{Throughput $T$ of different HARQ schemes versus the normalized buffer size $C$ for Gaussian signaling and baseband compression without adaptive storing (solid lines) and with adaptive storing with optimal $\eta$ (dashed lines) ($R=4$ bit/s/Hz, $\textrm{SNR}=10$ dB, and $N_{max}=10$).}
\label{fig:TvsC_Gau}
\efig

We start by considering Gaussian signaling and plot the throughput of the HARQ schemes under study versus the normalized buffer size $C$ in Fig. \ref{fig:TvsC_Gau} for $R=4$ bits/s/Hz, $N_{max}=10$, and $\textrm{SNR}=10$ dB.
HARQ-IR is seen, as expected, to outperform all other strategies, but its throughput gain depends strongly on the available buffer capacity $C$.
As for HARQ-CC, C\&S is observed to be preferred over S\&C, showing that the C\&S mechanism uses the receiver's memory more efficiently by storing the combined packet rather than the individual packets.
Moreover, the conventional Chase combiner that does not account for the impact of quantization is seen to be highly suboptimal in the regime of low $C$.
This performance loss is recovered by implementing adaptive storing, here shown with a value of $\eta$ obtained via numerical optimization for each value of $C$.
For example, for $C=5$ bit/s/Hz, the optimal value of $\eta$ was found to be equal to $1$.
Note that, with adaptive storing, the S\&C mechanism outperforms C\&S in the low regime of $C$, although this is an artifact of the simple adaptive storing policy considered here and could be fixed by implementing more sophisticated policies.

\bfig[t]
\bc
\centering \epsfig{file=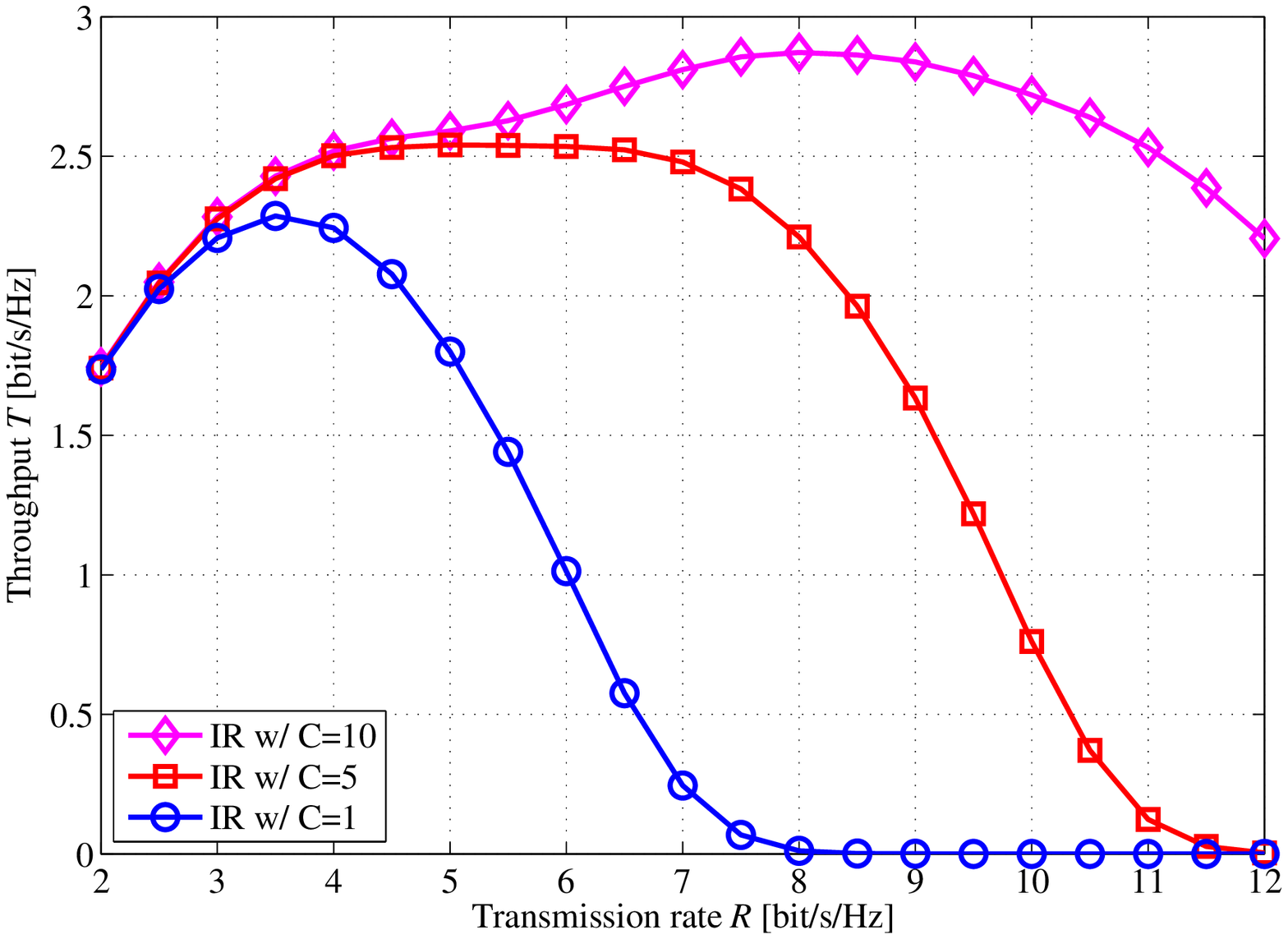, width=13cm, clip=}
\ec
\vspace*{-10mm}
\caption{Throughput $T$ of HARQ-IR versus the transmission rate $R$ with Gaussian signaling and baseband compression ($\textrm{SNR}=10$ dB and $N_{max}=10$).}
\label{fig:TvsR_IR}
\efig

In order to illustrate the importance of accounting for the available HARQ buffer capacity when designing the HARQ strategy, as done, e.g., with limited buffer rate matching in LTE \cite{sesia2009lte,Cheng},
we plot the throughput of HARQ-IR versus the transmission rate $R$ with $\textrm{SNR}=10$ dB, $N_{max}=10$, and different values of $C$ in Fig. \ref{fig:TvsR_IR},
It can be seen that the optimal value of $R$ depends significantly on the value of $C$, ranging from around $3.5$ bits/s/Hz for $C=1$ to $R=8$ bits/s/Hz for $C=10$ bits/s/Hz.
Further discussion on the advantages of adapting the HARQ strategy to the HARQ buffer via layered coding and blocklength optimization can be found below.

\bfig[t]
\bc
\centering \epsfig{file=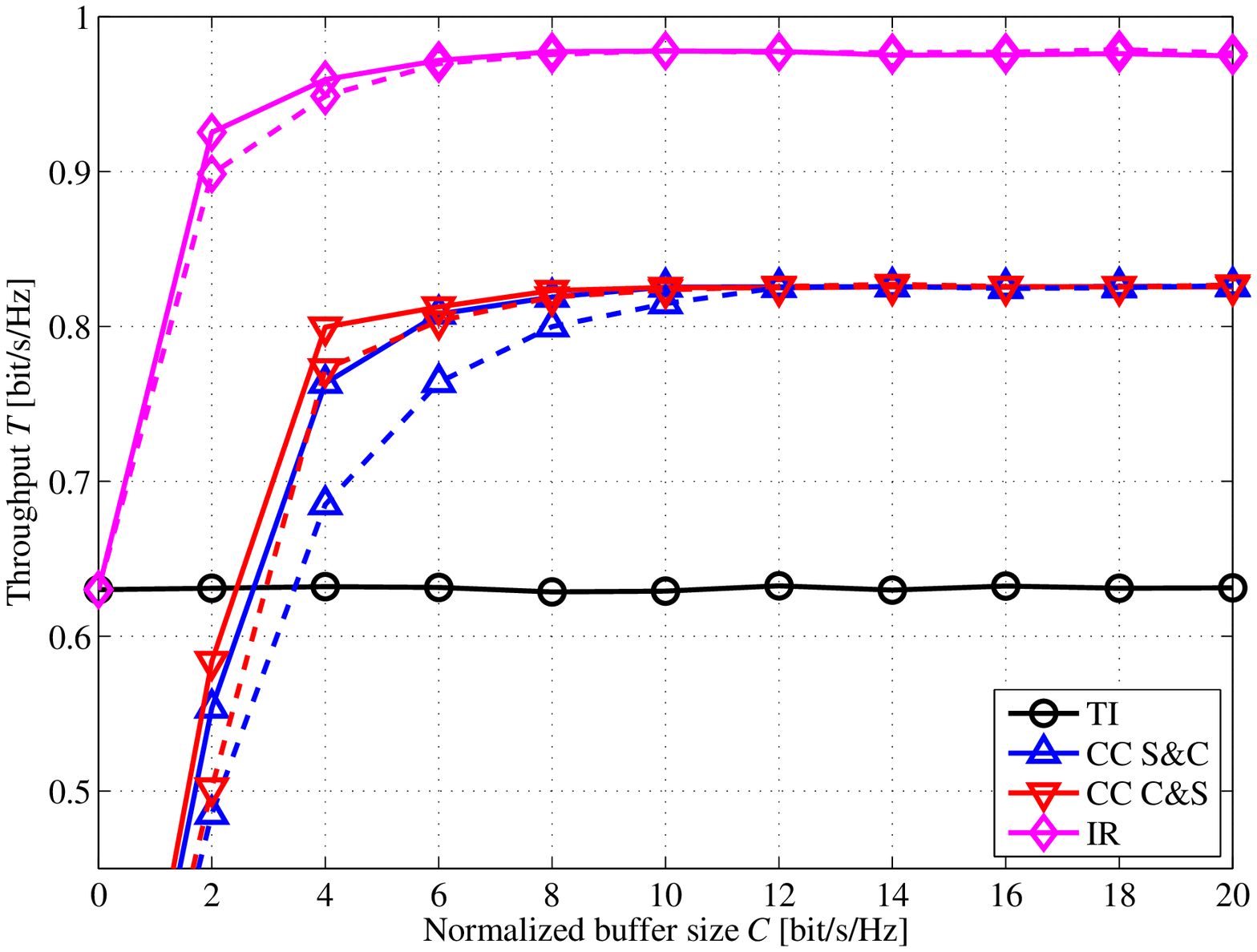, width=13cm, clip=}
\ec
\vspace*{-10mm}
\caption{Throughput $T$ of different HARQ schemes versus the normalized buffer size $C$ for BICM with $4$-QAM for baseband compression (solid lines) and LLR compression (dashed lines) ($M=4$, $R=1.6$ bit/s/Hz, $\textrm{SNR}=5$ dB, and $N_{max}=10$).}
\label{fig:TvsC_M4}
\efig

\bfig[t]
\bc
\centering \epsfig{file=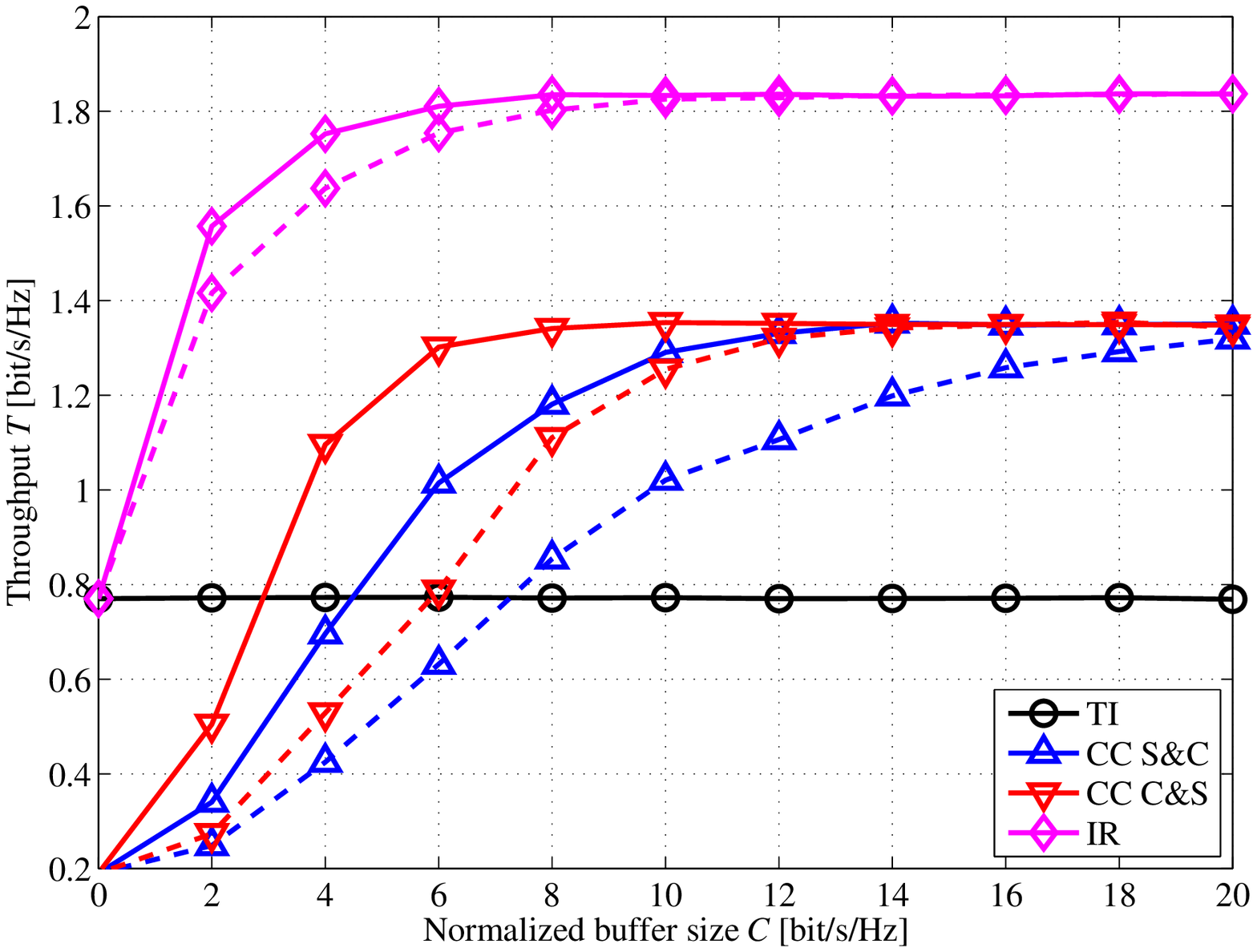, width=13cm, clip=}
\ec
\vspace*{-10mm}
\caption{Throughput $T$ of different HARQ schemes versus the normalized buffer size $C$ for BICM with $16$-QAM for baseband compression (solid lines) and LLR compression (dashed lines) ($M=16$, $R=3.4$ bit/s/Hz, $\textrm{SNR}=10$ dB, and $N_{max}=10$).}
\label{fig:TvsC_M16}
\efig

We then turn to the performance with BICM under both baseband and LLR compression. Fig. \ref{fig:TvsC_M4} and Fig. \ref{fig:TvsC_M16} show the throughput of different HARQ schemes under both compression strategies with $N_{max}=10$ for two modulation schemes.
Specifically, for Fig. \ref{fig:TvsC_M4}, we set the constellation to $4$-QAM, i.e., $M=4$, and the other parameters as $R=1.6$ bits/s/Hz and $\textrm{SNR}=5$ dB;
instead, for Fig. \ref{fig:TvsC_M16}, we set the constellation to $16$-QAM, i.e., $M=16$, with $R=3.4$ bits/s/Hz and $\textrm{SNR}=10$ dB.
Note that adaptive storing is not considered here in order to preserve the legibility of the figure but the performance with adaptive storing follows the same considerations as for Fig. \ref{fig:TvsC_Gau}.
It is seen that baseband compression is generally advantageous over the conventional LLR compression and that the relative gain is more pronounced for simpler HARQ strategies such as TI and CC.
This suggests that the use of a more sophisticated decoder, as in HARQ-IR, reduces the performance loss of a less effective compression strategy.
Moreover, by comparing Fig. \ref{fig:TvsC_M4} and Fig. \ref{fig:TvsC_M16}, we can see that the performance loss of LLR compression increases as the size of the constellation grows larger, particularly for simpler HARQ schemes.
This is due to the larger number of LLR  values that need to be compressed as the size of the constellation increases.

\bfig[t]
\bc
\centering \epsfig{file=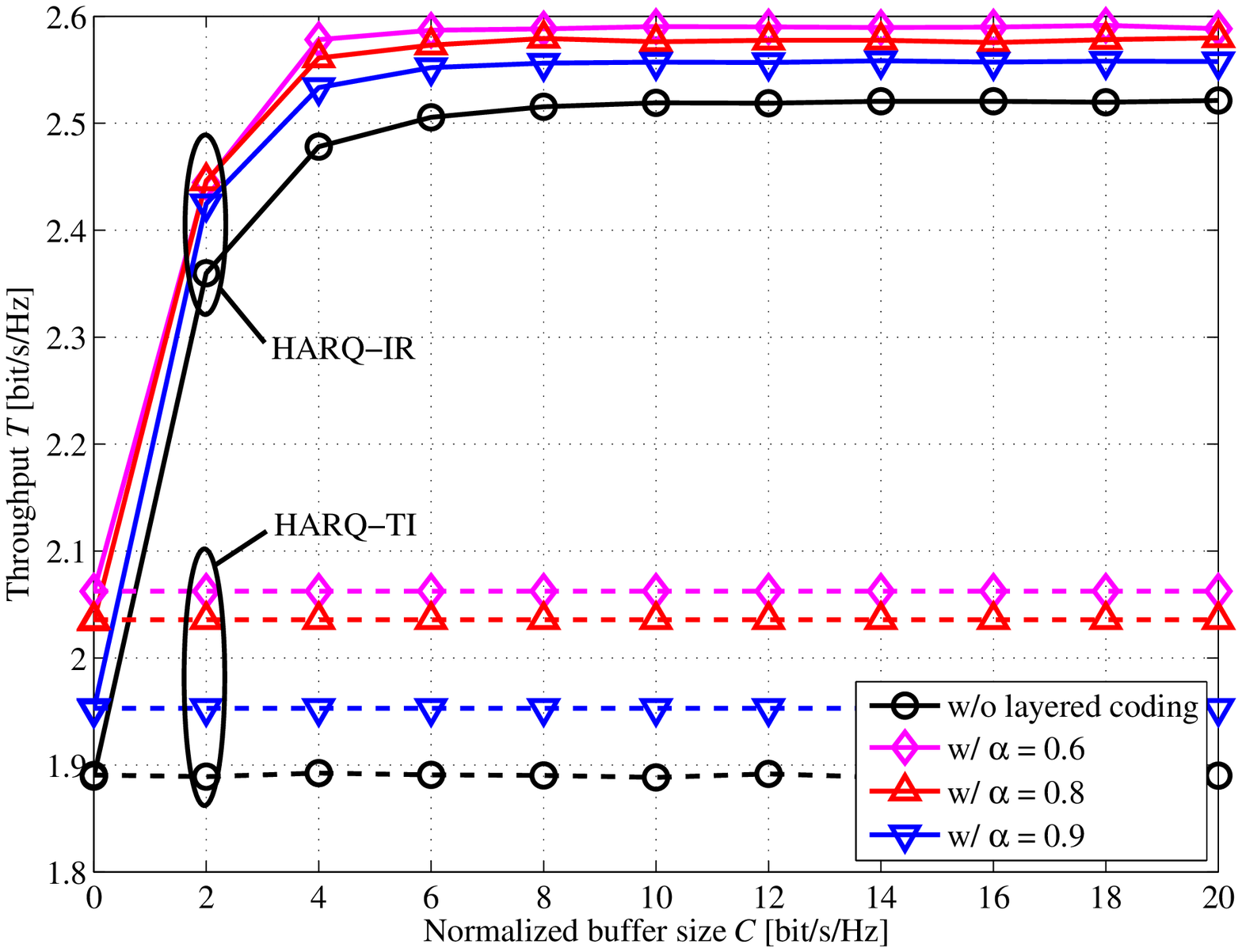, width=13cm, clip=}
\ec
\vspace*{-10mm}
\caption{Throughput $T$ of HARQ-TI and HARQ-IR versus the normalized buffer size $C$ for Gaussian signaling and baseband compression with layered coding with optimal $R_{1}$ ($R=4$ bit/s/Hz, $\textrm{SNR}=10$ dB, and $N_{max}=10$).}
\label{fig:TvsC_LC}
\efig

\bfig[t]
\bc
\centering \epsfig{file=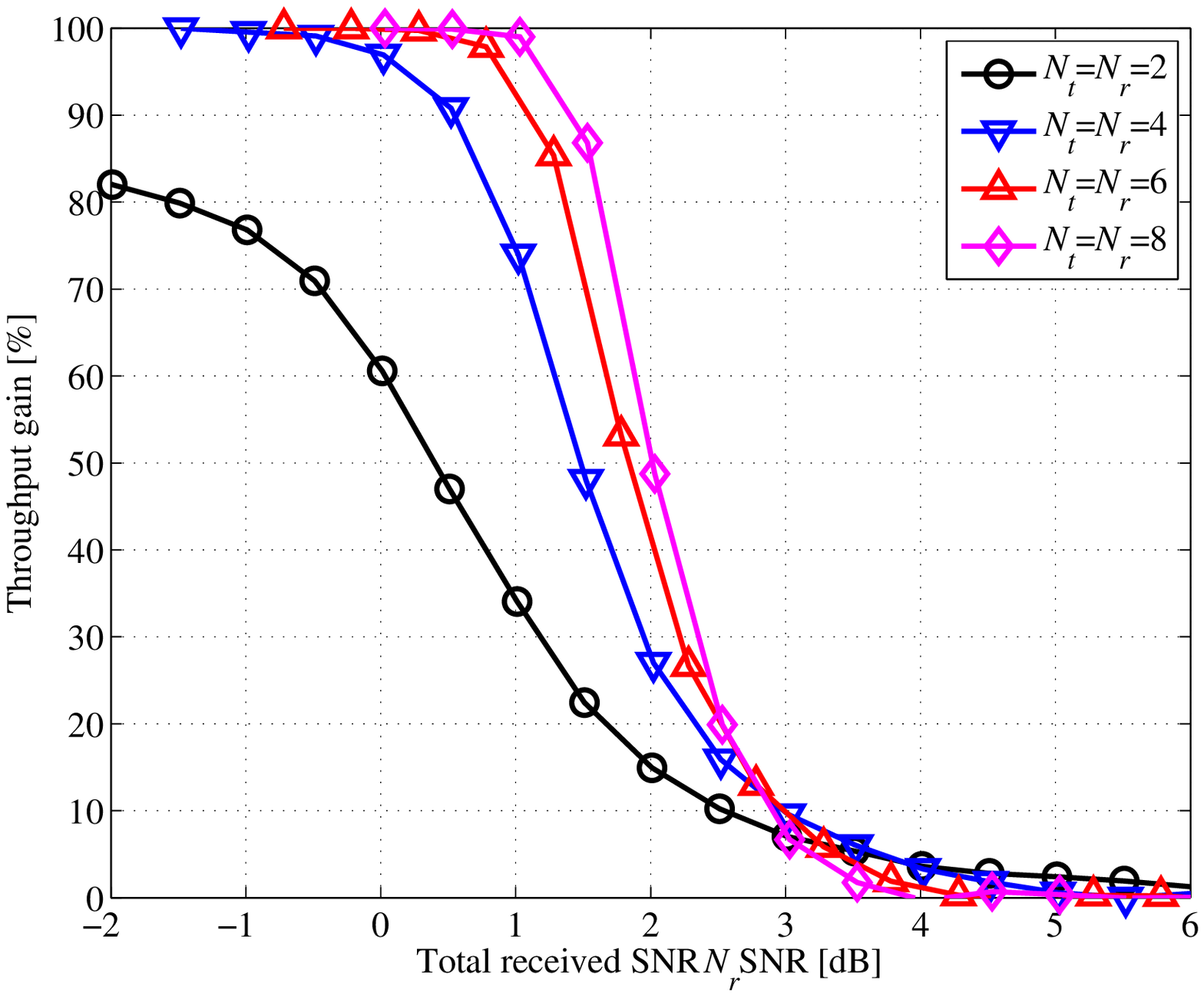, width=13cm, clip=}
\ec
\vspace*{-10mm}
\caption{Throughput gain of HARQ-IR versus the total received SNR $N_{r}\textrm{SNR}$ for Gaussian signaling and baseband compression ($R=5$ bit/s/Hz, $C=5$ bit/s/Hz, and $N_{max}=10$).}
\label{fig:TvsSNR_MIMO}
\efig

We now discuss the performance enhancement that can be obtained via two-level layered coding as presented in Sec. \ref{ch:LayeredCoding}.
To this end, Fig. \ref{fig:TvsC_LC} shows the throughput of HARQ-TI and HARQ-IR with $R=4$ bit/s/Hz, $\textrm{SNR}=10$ dB, and $N_{max}=10$.
The curves are derived by optimizing numerically over the value of the rate $R_{1}$ of the first layer with $0\leq R_{1} \leq R$, and we consider different values of the power splitting factor $\alpha$.
It is first observed that the throughput is quite sensitive to the choice of the power splitting factor $\alpha$.
Moreover, confirming the discussion in Sec. \ref{ch:LayeredCoding}, we see that the performance gain of layered coding is particularly pronounced in the regime of low $C$.
For instance, the throughput is increased by $9\%$ with layered coding at $C=0$ bit/s/Hz with $\alpha=0.6$, but only by $2\%$ for a sufficiently large $C$.

Next, we study the throughput performance of HARQ-IR for a MIMO link in Fig. \ref{fig:TvsSNR_MIMO} following the treatment in Sec. \ref{ch:MIMO}.
We plot the throughput gain of HARQ-IR versus the total received SNR for different numbers of transmit/receive antennas for $R=5$ bit/s/Hz, $C=5$ bit/s/Hz, and $N_{max}=10$.
The performance with the optimal transform coding matrix based on (\ref{eq:op_alpha}) is compared with a baseline solution in which $\mathbf{A}_{i,n}=k\mathbf{I}$, where $k$ is selected so as to satisfy the condition (\ref{eq:const_buffer}).
In Fig. \ref{fig:TvsSNR_MIMO}, the throughput gain is seen to be particularly significant as the number of antenna increases and in the regime of small received SNR.

\bfig [t]
\bc
\centering \epsfig{file=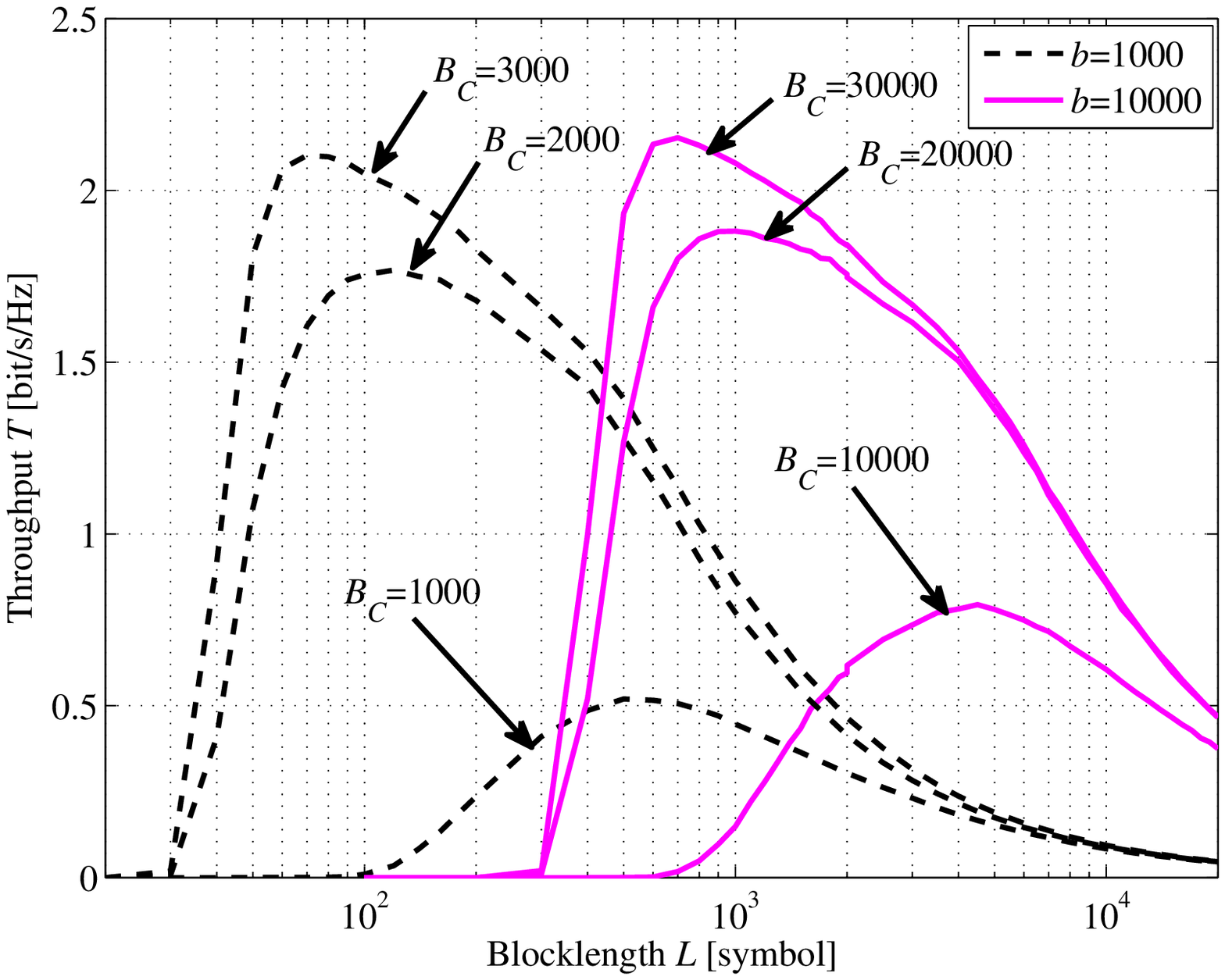, width=13cm,clip=}
\ec
\vspace*{-10mm}
\caption{Throughput $T$ of HARQ-IR versus the blocklength $L$ for Gaussian signaling and baseband compression ($\epsilon_{q}=10^{-4}$, $\textrm{SNR}=5$ dB, and $N_{max}=10$).}
\label{fig:TvsL_Bc}
\efig

\bfig[t]
\bc
\centering \epsfig{file=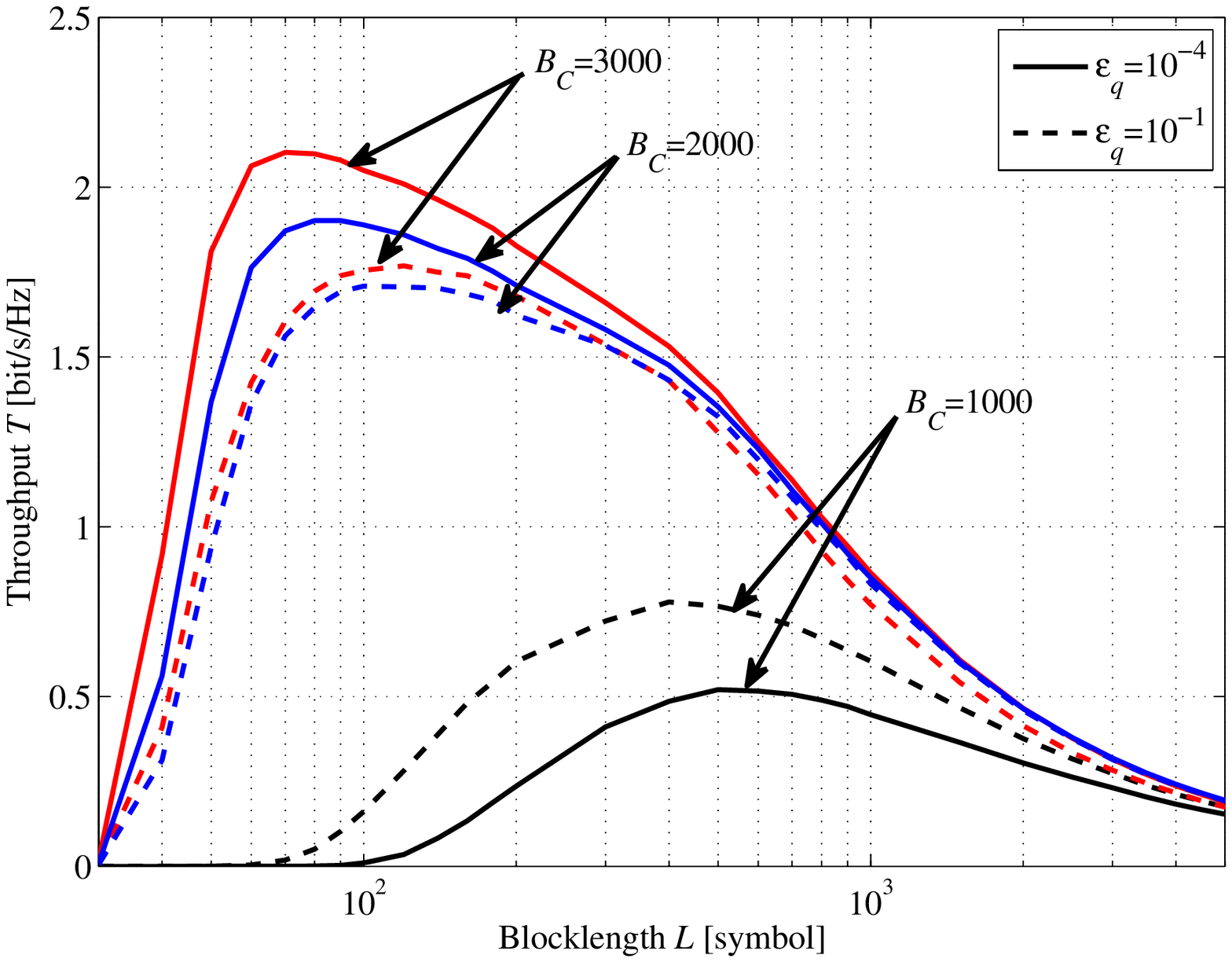, width=13cm,clip=}
\ec
\vspace*{-10mm}
\caption{Throughput $T$ of HARQ-IR versus the blocklength $L$ for Gaussian signaling and baseband compression ($b=1000$ bits, $\textrm{SNR}=5$ dB, and $N_{max}=10$).}
\label{fig:TvsL_ep}
\efig

We then discuss the impact of finite blocklength in the presence of a limited HARQ buffer as per the discussion in Sec. \ref{ch:F_Blocklength}.
Fig. \ref{fig:TvsL_Bc} shows the throughput $T$ versus the blocklength $L$ for different total buffer sizes $B_{C}$ with $\epsilon_{q}=10^{-4}$, $\textrm{SNR}=5$ dB, and $N_{max}=10$.
An increase in $B_{C}$ is seen to yield a significantly enhanced throughput for any value of the blocklength $L$, unless $L$ is large enough to overwhelm the limited HARQ buffer.
Moreover, a larger $B_{C}$ calls for a reduction in the blocklength $L$ in order to optimize the throughput.
This is because, with a larger HARQ buffer, more retransmissions can be accommodated and hence it is advantageous to transmit the first packet with a more aggressive rate $b/L$.
Finally, a smaller value of $b$, here $b=1000$, yields essentially the same throughput of a larger value, here $b=10000$, while entailing a smaller average delay.
For example, for the respective throughput maximizing values of $L$ and $B_{C}=30000$, we have the average delay (see \cite{Devassy}) $LE[N]=451$ with $b=1000$ bits and $LE[N]=4642$ with $b=10000$ bits.

Finally, we elaborate on the effect of the compression failure probability $\epsilon_{q}$ in Fig. \ref{fig:TvsL_ep}.
We set $b=1000$ bits, $\textrm{SNR}=5$ dB, and $N_{max}=10$.
As discussed in Sec. \ref{ch:F_Blocklength}, the choice of $\epsilon_{q}$ is one between a less significant back-off from the theoretical optimal distortion (large $\epsilon_{q}$) and a smaller probability of quantization failure (small $\epsilon_{q}$).
For small HARQ buffers, the quantization noise is large irrespective choice of $\epsilon_{q}$, and hence a small $\epsilon_{q}$, which minimizes the probability of dropping received packets due to quantization errors, is to be preferred.
Instead, for large HARQ buffers, the performance loss due to an excessive back-off from the optimal distortion is significant and then a larger value of $\epsilon_{q}$ is preferable.

\section{Conclusions} \label{ch:conclusion}
Motivated by the observation that, in modern wireless communication standards, such as LTE, the chip area occupied by the HARQ buffer is becoming increasingly significant, this work has taken an information-theoretic view of the problem of HARQ buffer management.
With reference to the questions asked in the introduction, our analysis has provided three important results.
(\textit{i}) We have quantified the performance advantage that can be accrued by more sophisticated HARQ schemes such as HARQ-IR as a function of the HARQ buffer size, demonstrating that the gains depend critically on the available buffer resources.
(\textit{ii}) We have shown that storing baseband samples is generally advantageous over the conventional strategy of storing LLRs, particularly for larger constellations.
Moreover, baseband compression enables sophisticated compression techniques to be implemented for multiple-antenna links, such as transform coding (see question (\textit{iv})).
This conclusion suggests that advanced compression mechanisms have the potential to dramatically reduce the necessary HARQ memory.
(\textit{iii}) We have investigated the potential benefits of buffer-aware transmission by considering layered modulation and the optimization of the transmission blocklength.
The results demonstrate that layered modulation is particularly advantageous in the presence of small HARQ buffers, and that smaller blocklengths, and hence a more aggressive transmission rate, are beneficial for larger HARQ buffers that can accommodate more received packets.

\section*{Appendix}
As discussed in Sec. \ref{ch:HARQ_BBcomp}, the HARQ-CC S\&C and HARQ-IR schemes operate by compressing all the received packets and allocating an equal fraction of the available memory to all compressed packets.
Therefore, a packet that has been already compressed to $LC/(n-1)$ bits at the $(n-1)$-th transmission needs to be recompressed at the $n$-th transmission (if unsuccessful) to a smaller number $LC/n$ of bits.
In this section, we explain how this can be accomplished by using successive refinement (or layered) coding. In so doing, we demonstrate that the equality (\ref{eq:Cn}) is valid also for recompressed packets.
Note that we discuss here the case of Gaussian signaling, but the treatment of BICM follows in a similar fashion.

Consider the compression of a received $i$-th packet as in (\ref{eq:system_model}).
The packet can be recompressed at most $N_{max}-i$ times since there are at most as many possible retransmissions in which the packet at hand can be reused by the decoder.
To enable this, if the $i$-th transmission is unsuccessful, the decoder compresses (\ref{eq:system_model}) by using a successive refinement code with $N_{max}-i$ layers,
which corresponds to the progressively less accurate compressions that are stored in subsequent retransmissions. Specifically, for each packet $i$,
we have the descriptions $\hat{Y}_{i,n}=Y_{i}+Q_{i,n}$ in (\ref{eq:comp_packet}) for $n=i,i+1,\dots,N_{max}-1$.
The corresponding quantization noise variances $\sigma_{i,n}^{2}$ are increasing in $n$, since the allocated memory becomes smaller in later transmissions, i.e., we have the inequalities
\be \label{eq:ineq_sr}
\sigma_{i,i}^{2}\leq\sigma_{i,i+1}^{2}\leq\cdots\leq\sigma_{i,N_{max}-1}^{2}.
\ee

As summarized in Fig. \ref{fig:App} for each packet $i$, the decoder produces the ``base layer" description $\hat{Y}_{i,N_{max}-1}$, which has the largest quantization noise variance $\sigma_{i,N_{max}-1}^{2}$,
and the ``refinement layers" $\hat{Y}_{i,N_{max}-2},~\hat{Y}_{i,N_{max}-3},\dots,~\hat{Y}_{i,i}$ with progressively smaller quantization noise variances as per (\ref{eq:ineq_sr}).
At the $j$-th retransmission, with $j=i,\dots,N_{max}-1$, only the descriptions $\hat{Y}_{i,N_{max}-1},~\hat{Y}_{i,N_{max}-2},\dots,~\hat{Y}_{i,j}$ are stored
and the higher refinement layers are discarded.

Based on the discussion above, we can write the quantization noises as
\be \label{eq:Qnoise}
Q_{i,n}=\sum_{j=i}^{n}\Delta Q_{i,j},
\ee
where the variables $\Delta Q_{i,j}\sim CN(0,\sigma_{i,j}^{2}-\sigma_{i,j-1}^{2})$ are independent and represent the increase in quantization noise variance in going from the $(j-1)$-th description to the $j$-th (we set $\sigma_{i,i-1}^{2}=0$).
This shows that the Markov chain ${Y}_{i}-\hat{{Y}}_{i,i}-\hat{{Y}}_{i,i+1}\cdots-\hat{{Y}}_{i,N_{max}-1}$ holds.
Moreover, using standard information-theoretic results on the performance of successive refinement (see, e.g., \cite[Ch.~13]{Gamal}), we obtain that the number $R_{i,n}$ of bits per symbol needed to store the $n$-th description is given by
\be
R_{i,n}=I(Y_{i};Y_{i}+Q_{i,n}|Y_{i}+Q_{i,n+1})
\ee
for $n=i,\dots,N_{max}-2$ and $R_{i,N_{max}-1}=I(Y_{i};Y_{i}+Q_{i,N_{max}-1})$.
The overall number of bits per symbol that need to be stored at the $n$-th transmission (if unsuccessful) is hence given by
\be
\sum_{j=n}^{N_{max}-1}R_{i,j}=I(Y_{i};Y_{i}+Q_{i,n}),
\ee
which can be seen by recalling the definition of condition mutual information\footnote{$I(A;B|C)=h(A|C)-h(A|C,B)$ for continuous jointly distributed variables $A$, $B$, $C$, where $h(A)$ is the differential entropy of variable $A$.}
and noticing that, because of the mentioned Markov chain relationship, we have $h(Y_{i}|Y_{i}+Q_{i,n},Y_{i}+Q_{i,n+1})=h(Y_{i}|Y_{i}+Q_{i,n})$.
We conclude that the equality (\ref{eq:Cn}) holds also for recompressed packets.

%
%
%
%
%




\end{document}